\documentclass[]{interact}
\usepackage{float}
\usepackage{comment}
\usepackage[table]{xcolor}
\usepackage[euler]{textgreek}
\usepackage{epstopdf}
\usepackage[caption=false]{subfig}
\usepackage[natbibapa,nodoi]{apacite}
\theoremstyle{plain}

\theoremstyle{definition}

\theoremstyle{remark}

\begin{document}

\title{
Rigid and Aeroelastic Analysis of Wind Induced Flow Behavior Around Unscaled Tall Buildings using Numerical Techniques}
\author{
\name{V.A.U. De Alwis\textsuperscript{a*}\thanks{CONTACT V.A.U. De Alwis.
Email:*arjunadealwis@eng.pdn.ac.lk} , P.H.T.D. Weerarathne\textsuperscript{a} and K.K. Wijesundara\textsuperscript{b}}
\affil{\textsuperscript{a}Department of Mechanical Engineering, University of Peradeniya, Sri Lanka; \textsuperscript{b}Department of Civil Engineering, University of Peradeniya, Sri Lanka} }
\maketitle

\begin{abstract}
The study focuses on the wind loads acting on the standard Commonwealth Advisory Aeronautical Council (CAARC) building model. Numerical modeling has been used instead of employing wind tunnel experiments due to the requirement of high physical space. This is mainly due to the superiority of CFD codes included in turbulence models such as RANS and more precise LES. However, the shortcoming arising through basing the result on numerical investigations is compensated by direct comparison to experimental data available in the literature. The study extends beyond the typical rigid building structure giving account to the minor deformations resulting from the impact of the wind force on an aeroelastic structure through coupled two-way fluid-structure interaction while designing the building to have its natural fundamental frequency of 0.2 Hz. An improved Smagorinsky-Lily model was used for all LES models and the results were compared with available wind tunnel experimental data.  It was found that coupled LES analysis provided a better correlation to wind tunnel data compared to rigid fluid domains with respect to pressure distribution and wake formation despite the additional computation time taken.  However, all these readings were dependent on the boundary conditions which need to be estimated concisely fur future use.

\end{abstract}

\phantom{\textcolor{blue}{\cite{hirt},\cite{zeng20}
\cite{hirt}
\cite{hirt75}
\cite{hans82}
\cite{hans86}
\cite{sum86}
\cite{blocken}
\cite{yuan}
\cite{huang}
\cite{koli17}
\cite{mou}
\cite{koli18}
\cite{dagnew}
\cite{abo}
\cite{akbay}
\cite{jansen}
\cite{chima}
\cite{chen}
\cite{lau93}
\cite{tsu97}
}}
\newpage
\section{Introduction}
The popularity of tall buildings is growing due to the population growth and the limited usable land in the major cities around the world. The geometrical complexity and the maximum possible height of the buildings have already reached a point where traditional methods for analyzing wind-induced responses are obsolete due to the cost attached with it and the limited accuracy. Moreover, \cite{zeng20} concludes that conventional wind tunnel test techniques (including HFBB, SMPSS and aeroelastic test techniques) and advanced wind tunnel test techniques (including forced vibration test techniques and hybrid aerodynamic-pressure/force balance test techniques) are not fully capable of predicting wind-induced responses. As the height of the building increases, it becomes more susceptible to wind loads. Therefore, Computational Fluid Dynamics (CFD) provides a viable alternative for expensive wind tunnel tests, at least in the preliminary design stage in the design to determine wind-induced loads on high-rise buildings. But, even with the modern-day computational power, it is impractical to obtain Direct Numerical Solutions (DNS) to the Navier Stokes equations which represent the problem in differential equations. As alternatives, Reynolds Averaged Navier Stokes (RANS) and Large Eddy Simulation (LES) models are widely used for the evaluation of wind effects on structures. RANS mainly relies on statistical approaches that display a time-mean representation of the fluid flow. In contrast, LES is a multi-scale computational modeling approach that offers a more comprehensive way of capturing fluctuating turbulent flow by resolving eddies formed due to turbulence. However, most of the studies available have used the RANS approach instead of more accurate LES due to the high computational power required for the latter.\par

Since the early 1980s, numerical studies of building aerodynamics have been conducted. \cite{hirt}, one of the first papers dedicated to numerical simulation of the wind action over bluff bodies, predicted that there are many possibilities and considerable potential for achieving useful results in the future due to hardware and software development of the computer resources. In that particular study, a finite difference model is presented and applied to some simple applications on bluff body aerodynamics through a solution algorithm developed for transient fluid flows and presented in \cite{hirt75}. There, the low numerical resolution of the grid in the vicinity of the bluff bodies in basically every approximation method was highlighted as one of the causes that contributes to the less accurate results.\par
The very beginning of investigating aerodynamics around buildings goes back to 1982. \cite{hans82}, \cite{hans86} simulated wind flow around building models with different configurations and compared to wind tunnel tests to confirm the reliability by the means of experiments. The importance of using a 3D model instead of using 2D planes to evaluate the simulation is emphasized in the above papers. Furthermore,\cite{sum86} discusses the inability to accurately solve the wake region of the time-averaged flow around a bluff surface-mounted obstacle in a boundary-layer wind tunnel. It is identified that, with the approximate treatment of solid boundaries, simulations cannot serve to accurately determine the structural loading of buildings. The standard \emph{k}- \textepsilon\ model was revised and a turbulence model was proposed by \cite{lau93} called the LK model. \cite{mura98} presented the further developed MMK turbulence model addressing the issues of the LK model as commented in \cite{tsu97}. In a more recent study, \cite{koli18} compares 7 different turbulence models with available wind tunnel test data and concludes that most of the turbulence models fail to accurately predict pressure distribution of the leeward and side faces of the body due to the flow separation. Therefore, the wind engineering community moved towards LES to analyse wind loads and induced vibrations of building models due to complex and unsteady turbulent flows, containing wake regions.
\par
\cite{mura89}, \cite{mura90}, \cite{mura93}, \cite{mura97} discuss turbulent modeling and applications of turbulence models to predict the wind-induced vibration of building models. Murakami (1993) concludes that, even though LES requires higher computational power, it provides higher accuracy in results. The reason being, in LES, the time-dependent inflow boundary is responsible for a more accurate representation of oncoming fluid flow properties.  Hence \cite{dagnew} and \cite{swad} recommended LES as a complementary tool for wind load evaluation. \cite{huang} performed a comprehensive analysis of the CAARC building model employing numerical simulation where the building is considered as rigid and aerodynamic coefficients and flow patterns around the building were predicted through a CFD commercial software. Despite the fact that the dynamic properties such as the natural fundamental frequency of the building are discussed in the studies above, the coupled interaction of the building configuration with the fluid domain is not taken into consideration. 
\par

So far, the Computational Wind Engineering (CWE) community has done a number of numerical simulations based on wind effects over different building configurations. While some of the numerical solutions of the above work focus on problems such as interference effects between buildings (\cite{sum86}), wind-driven environmental conditions and pedestrian comfort (\cite{mochida}; \cite{ramponi} ; \cite{stath}; \cite{yuan}), others shed light on the wind interactions with structures (\cite{huang}; \cite{braun}; \cite{koli17}; \cite{mou}; \cite{chen}). It is observed that the majority of the studies focus on numerical analysis on rigid building models rather than aeroelastic building models. \cite{bath} suggests a developed scheme, which complies with engineering practice for fluid-structure interaction problems and further discusses the potential for an ideal scheme. \cite{rugo} employs Lyapunov characteristic exponents to determine the long-term dynamic stability of fluid-structure interaction systems, allowing them to distinguish between chaotic and regular system behavior. \cite{abo} used Von Karman’s spectrum and Davenport’s Coherency function to digitally create turbulent velocity vectors, since LES in Ansys FLUENT package needed previous knowledge of time-dependent velocity vectors of the atmospheric boundary layer. There, the simulations were run to calculate the wind-induced response of tall buildings that exhibit coupled modal behavior.\par

This study aims to counter most of the drawbacks that have not been addressed in the previous studies described above so that to build a numerical simulation that perfectly aligns with an experimental setup. The study comprises several novelties. Mainly, the use of a partitioned coupling approach between the building and wind domain has been the main aspect highlighted in this study which has been barely touched in the literature in the context of application for wind loads. Furthermore, even the studies having the closest resemblance to the coupling code have not considered the frequency modes of the original building which is taken into consideration in this study. Moreover, an efficient block-based mesh method was introduced in order to generate a finer mesh near the boundary region which gets coarser when moving away from the building leading to optimized computational performance. Furthermore, rather than using a scaled version of the actual setup, the model was analyzed considering full-scale dimensions to avoid inaccuracies that would cause by using different scaling theories. Finally, an intra-study is conducted comparing three arbitrary simulation cases using the same grid to compare the effect of turbulence and coupling methods for pressure approximation.\par

 For this purpose, three distinct analysis methods are used to analyze the wind-induced response, along and across the CAARC building and the results are compared with experimental data obtained at six research centers; The City University (CU), University of Bristol (BU), Monash University (MU), National Aeronautical Establishment (NAE), National Physical Laboratory (NPL), and University of Western Ontario (UWO) \citep{mel}, and also the experimental readings obtained from Goliger and Milford (1988). The comparison is focused to analyze the dependence of stiffness of the building, and the dependence on whether the fluid flow analysis is RANS or LES has a significant impact on the result. Initially, in Case 1 and Case 2, a comparison is drawn between the dependence of different numerical models by comparing the results generated for \emph{k}- \textepsilon\ turbulence model and the Smagorinsky-Lilly subgrid-scale model respectively for a rigid fluid model. Finally, the aeroelastic nature of the building was also considered in Case 3 by implementing a two-way coupled simulation setup between fluid and structural domains where the same Smagorinsky-Lilly subgrid-scale model was used to estimate turbulence.\par

The RANS model, which is a way of resolving turbulent flows, solves time-averaged Navier Stokes equations in a steady-state, whereas the LES model predicts wind-driven reactions by filtering flow variables in order to resolve large scales of motion and represent just the influence of small ones. The LES model with the aeroelastic building configuration is expected to supersede the LES model with the rigid building configuration since it emulates the natural phenomena through two-way fluid-structure interactions.


\section{Theory for turbulence models}
\subsection{  {K}- \textepsilon\ model}
According to \cite{launder}, for low \emph{Re} formulation damping functions are applied to the model coefficients in the turbulence equations. As by using these model coefficients the mesh can be brought down to viscous sublayer (\emph{y}\textsuperscript{+}\(<\)5), \emph{k}-\textomega\ SST model is preferred for low \emph{Re} (Reynolds) number applications. Therefore, in general the \emph{k}-\textepsilon\  model is preferred for high \emph{Re} number applications  when the wall \emph{y}\textsuperscript{+} is not within the buffer region (5\(<\)\emph{y}\textsuperscript{+}\(<\)30).\par
The \emph{Re} number for the flow in the fluid domain considered was measured at a value of 1.36×10\textsuperscript{4} and due to this \emph{k}-\textepsilon\ model was used in this study as a RANS turbulence model. Further, various literature have shown high accuracy results for wind flows lying around these \emph{Re} number ranges once  \emph{k}-\textepsilon\ model is used \citep{huang}. The theory associated with the standard \emph{k}-\textepsilon\ model is briefly described as follows. However, it is highly recommended to refer FLUENT theory guide \citep{ansth}. \par
 
The distinction between different RANS turbulent models is made with respect to the Reynolds stress term in the Navier Stokes Equations customized for RANS models as given in Equation \ref{eqn1}.

\begin{equation} \label{eqn1}
\frac{\partial(\rho U)}{\partial t}+\nabla \cdot(\rho UU) = -\nabla p+ \nabla .[\mu (\nabla U+(\nabla U)^T)] + \rho g -\nabla (\frac{2}{3} \mu (\nabla . U)) - \textcolor{red}{\nabla . (\overline{\rho U'U'}}) 
\end{equation}

All these symbols are used according to their standard definitions \citep{ansth}.  
The Reynolds stress term is denoted by red and in order to evaluate this commonly the Boussinesq hypothesis is used. The hypothesis can be stated as in Equation \ref{eqn2}.

\begin{equation} \label{eqn2}
-\overline{\rho U'U'} = \mu_t (\nabla U + (\nabla U)^T) - \frac{2}{3}\rho k I - \frac{2}{3} (\nabla . U) I
\end{equation}

Here , \emph{\textmu \textsubscript{t}} denotes the eddy viscosity and \emph{I} is the standard symbol used for turbulent intensity.
It is obvious according to the equation that in order to calculate the Reynolds stress term the eddy viscosity needs to be estimated. Different methods have been used in the literature in order to calculate this value.The most commonly used approach for eddy viscosity formulation is by using transport equations for \emph{k} and \textepsilon\ terms and thus it could be solved according to Equation \ref{eqn3}.

\begin{equation} \label{eqn3}
\mu_t = C_\mu \frac{\rho k^2}{\epsilon}
\end{equation}

The transport equation for \emph{k} is normally the same for all RNG, realizable and standard \emph{k}-\textepsilon\ models and it is written as in Equation \ref{eqn4}.(The four terms at the end are used for sources and sinks.)

\begin{equation} \label{eqn4}
\frac{\partial(\rho \epsilon)}{\partial t} + \nabla . (\rho U k) = \nabla . [(\mu+ \frac{\mu_t}{\sigma_k})\nabla k] + P_k + P_b - \rho \epsilon + S_k
\end{equation}

However, this is not the same for the \textepsilon\ terms. The calculation for the \textepsilon\ term is done according to the following Equation \ref{eqn5}.

\begin{equation} \label{eqn5}
\frac{\partial(\rho \epsilon)}{\partial t} + \nabla . (\rho U \epsilon) = \nabla . [(\mu+ \frac{\mu_t}{\sigma_\epsilon})\nabla \epsilon] + C_1 \frac{\epsilon}{k}(P_k+C_3P_b)- C_2\rho \frac{\epsilon^2}{k} + S_\epsilon
\end{equation}
Where \emph{C\textsubscript{1}}, \emph{C\textsubscript{2}}, \emph{C\textsubscript{3}} denote the model coefficients for the turbulence and their values vary between each of those models chosen among RNG, realizable and  \emph{k}-\textepsilon\  models.The model coefficients used in the FLUENT code can be found in \cite{launder}. In each of these coefficients, separate damping functions are included within to account for the damping that occurs in the viscous sublayer. Due to this the \emph{k}-\textepsilon\ model has the capability to solve the turbulence dissipation upto the closest cell layer from the building wall making the results accurate. Far away from the wall the damping function approaches unity returning the high Reynolds number formulation of the \emph{k}-\textepsilon\ model. Since the results of \emph{k}-\textepsilon\ models are much dependent on these functions under low Reynolds number formulation, the model is generally preferred for high \emph{Re} applications where separation and reattachment are not abundant.\par
However, even if the turbulence dissipation is calculated accurately in order to determine the forces, viscous shear stress should also be taken into consideration. In order to estimate this correctly, wall functions are used in FLUENT. The software uses two types of wall functions which are standard and enhanced to calculate the viscosity. The use of the function depends on the \emph{y}\textsuperscript{+} values in the mesh region, and if the \emph{y}\textsuperscript{+} does not lie between 5 and 30 the standard wall functions would be recommended. This is because the linear estimate of viscosity falls in line with the viscosity predicted by Direct numerical simulations except for the buffer region where they deviate vastly from each other. \par
Upon generating the results for the RANS calculation, it was observed that the minimum \emph{y}\textsuperscript{+} recorded along the boundary of the building wall was 30, which means none of the cells is within the buffer region and all of them would lie in the log law region (\emph{y}\textsuperscript{+}\(>\)30). However, it is said that the best practice is to allocate the closest cells to the walls to the viscous sublayer due to different inefficiencies in the log law model that occur when there are adverse pressure gradients and strong curvature in the wall which is not seen in the model discussed in the study. Furthermore, due to the high number of elements that have already been established for this model it was impractical to reduce the \emph{y}\textsuperscript{+} further by decreasing the cell size based on the available computational resources. Moreover, as an unscaled model was used and the closest mesh block had a size of 0.25 m by employing an inflation layer to reduce the \emph{y}\textsuperscript{+} to the viscous sublayer would mean an abrupt increase in the aspect ratio of those cells which would reduce the mesh quality and fail to deliver the purpose expected by having a structured cartesian mesh. 

\subsection{Smagorinsky Lilly model}

Similar to the RANS turbulence models, subgrid-scale models in FLUENT also use the Boussinesq hypothesis to compute the subgrid-scale turbulent stresses using an eddy viscosity approach. The equations used to calculate the subgrid-scale turbulent stress and the strain tensor are shown in Equations \ref{eqn6} and \ref{eqn7} and all of them use the standard notations used in \cite{ansth}.

\begin{equation} \label{eqn6}
\tau_{sgs} = 2\rho \upsilon_{sgs} S^*_{ij}- \frac{2}{3}k_{sgs} \delta_{ij}
\end{equation}

\begin{equation} \label{eqn7}
S^*_{ij} = \frac{1}{2} (\frac{\partial U_i}{\partial x_j}+\frac{\partial U_j}{\partial x_i}-\frac{\partial U_k}{3\partial x_k}\delta_{ij})
\end{equation}

All the subgrid-scale models available today differ by the turbulent viscosity term in Equation 8. The original Smagorinsky-lilly model was developed in the 1960s and due to its defects in giving accurate predictions, FLUENT uses a modified set of equations. The basic equations in developing the subgrid-scale viscosity is given in Equation \ref{eqn8}.

\begin{equation} \label{eqn8}
\upsilon_{sgs}= l^2_0 * \sqrt{2S_{ij}S_{ij}}
\end{equation}

Where $l_0$ is the subgrid length scale which is similar to the integral length scale used in the RANS equations. Although the strain rate can be calculated from resolved eddies, the subgrid length scale is difficult to be determined directly. The original Smagorinsky Lilly model uses an estimation of  $C_s$Δ as the subgrid length scale where a value of 0.17 is used for the Smagorinsky coefficient ($C_s$) and Δ gives the cell size. But this coefficient is only accurate for a homogeneous isotropic turbulence scenario and would be inaccurate if used for a flow near walls due to the underprediction of shear. Furthermore, the original model would result in non-trivial viscosity modeled near the wall regions making the whole fluid domain filled with homogenous eddies. Due to this, a novel parameter called the mixing length is introduced which is expressed as in Equation \ref{eqn9}.

\begin{equation} \label{eqn9}
l_m=\kappa y
\end{equation}

Where \emph{\textkappa} is the Von Karman constant and \emph{y} is the distance from the wall. Due to this the eddy size would decrease when moving towards the wall. As the subgrid scale length scale cannot be greater than the mixing scale it could be expressed as in Equation \ref{eqn10} \citep{ansth}.

\begin{equation} \label{eqn10}
L_s=min(\kappa d , C_s \Delta)
\end{equation}

Furthermore, deviating from the original model $C_s$ is also used as 0.1 in FLUENT as it has been able to give accurate results for a wide range of simulations. However, this model is also valid outside the buffer region and different other adjustments such as the Van driest approximation have to be used in order to create successful numerical results if \emph{y}\textsuperscript{+} lies within the buffer region.   

\section{Numerical Simulation Setup}  
\subsection{Computational domain and mesh arrangement for fluid domain}

For this study, the geometric model was selected according to standard full-scale dimensions of CAARC building. The building was modeled as a 30 m×45 m×180 m cavity made in the fluid domain considered using the ANSYS Design modeler modeling package. As for the first two cases out of the three, the aeroelastic nature of the building is not taken into consideration; a mere modeling of a cavity together with a static mesh is sufficient.\par
\begin{figure}[h]
\centering
\subfloat[Side view of the fluid domain]{%
\resizebox*{8cm}{!}{\includegraphics{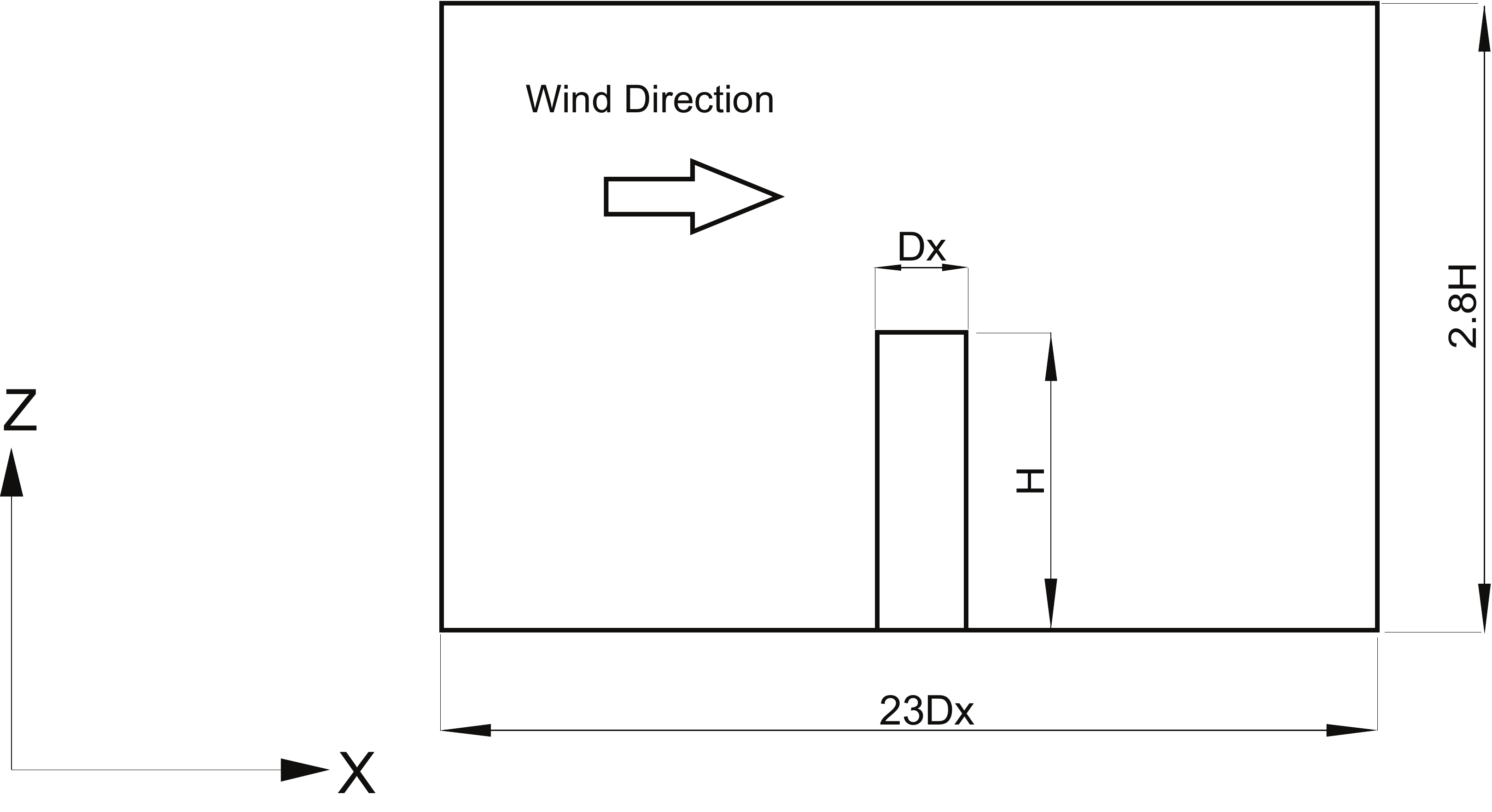}}}\hspace{5pt}
\subfloat[Top view of the fluid domain]{%
\resizebox*{8cm}{!}{\includegraphics{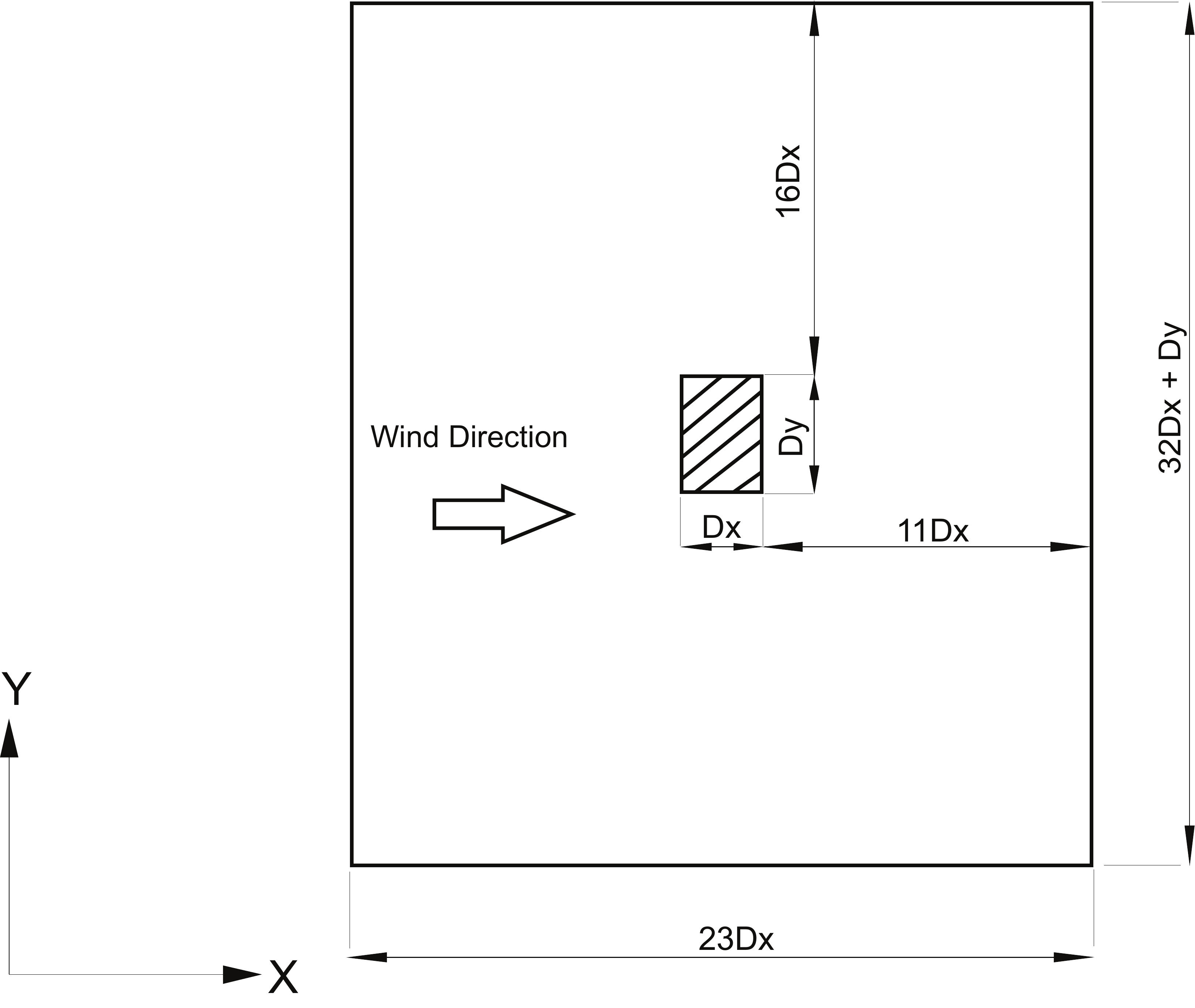}}}
\caption{Dimensions of the computational fluid domain} \label{views}
\end{figure}

In deciding the dimensions of the fluid domain to be considered for the flow around bluff bodies, special attention should be given to the obstacle effect that arises due to inflow and outflow conditions \citep{mura98}. This is also to allow the boundary layer to develop fully so that the properties within the vicinity of the building are captured efficiently. After considering different sets of guidelines, the dimensions of the domain were concluded to be as in Figure \ref{views}, where \emph{H,Dx, Dy} are building dimensions and \emph{H} = 180 m, \emph{Dx} = 30 m and \emph{Dy} = 45 m.\citep{koli18}. 

\begin{figure}[H]
\centering
{\includegraphics[scale=0.25]{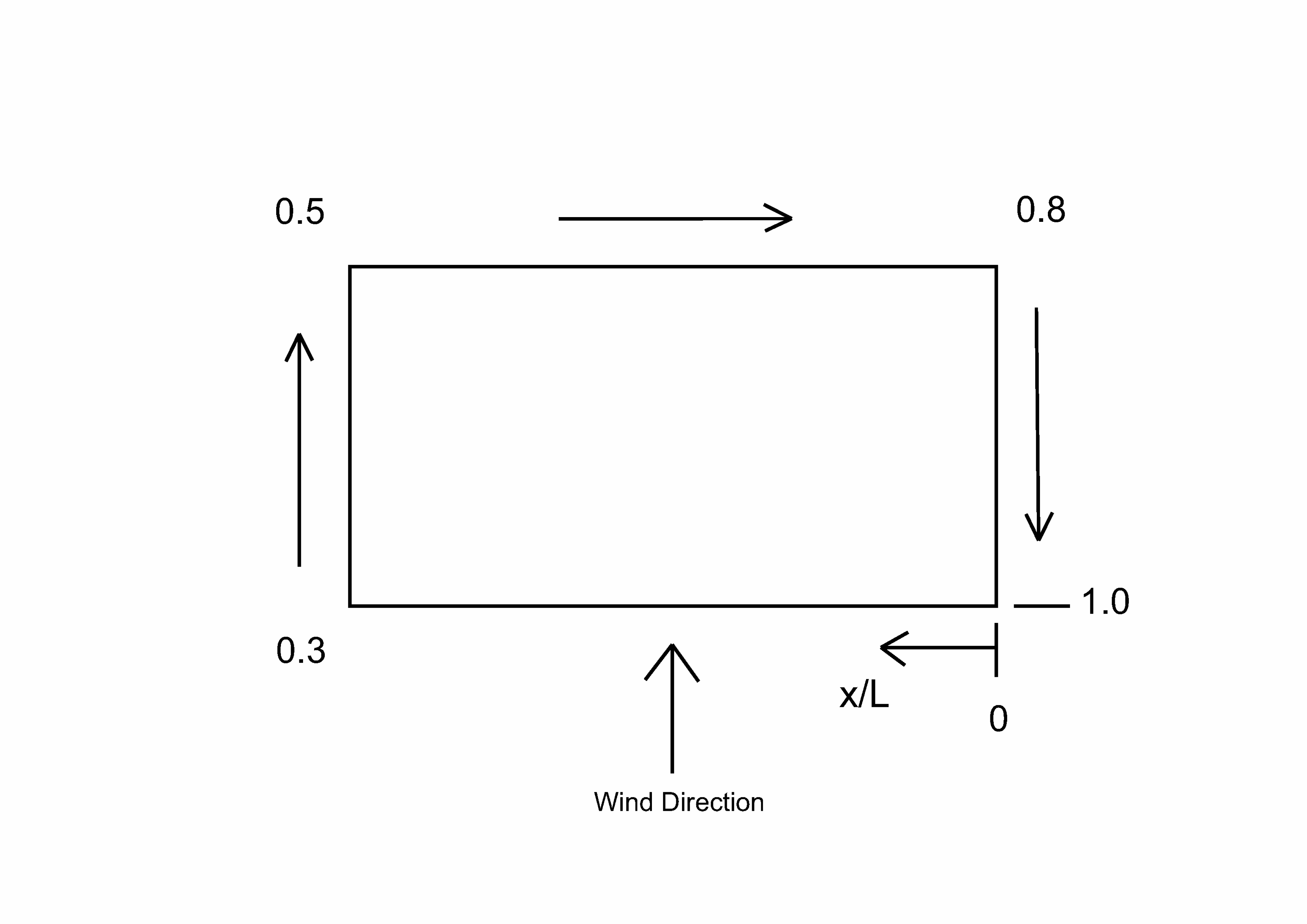}}\hspace{5pt}
\caption{Plot of x/L along the faces of the building.} \label{xL}
\end{figure}

As the cavity is confined to the middle, the interference of the flow pattern at the boundaries would hardly affect the results captured around the building. Furthermore, both along-wind and across-wind pressure distributions are captured in this analysis by plotting the results along the wall boundary at four sides as shown in Figure \ref{xL}. A parameter called (\emph{x}/\emph{L}) is introduced along the perimeter and consequently, the following values are established in the corners of the building.

Although the computational domain seems to be straightforward, the generation of the mesh seems to be intricated due to the compromise that has to be made between accuracy and computational efficiency based on the available resources. Besides, the mesh size has to be large enough so that the computational memory is sufficient enough to solve the computational matrix and should be small enough to capture the maximum amount of turbulent kinetic energy. According to \cite{sum86}, the grid should be fine enough to hold a minimum of 80 percent of turbulent kinetic energy.\par

\begin{figure}[H]
\centering
{\includegraphics[scale=0.2]{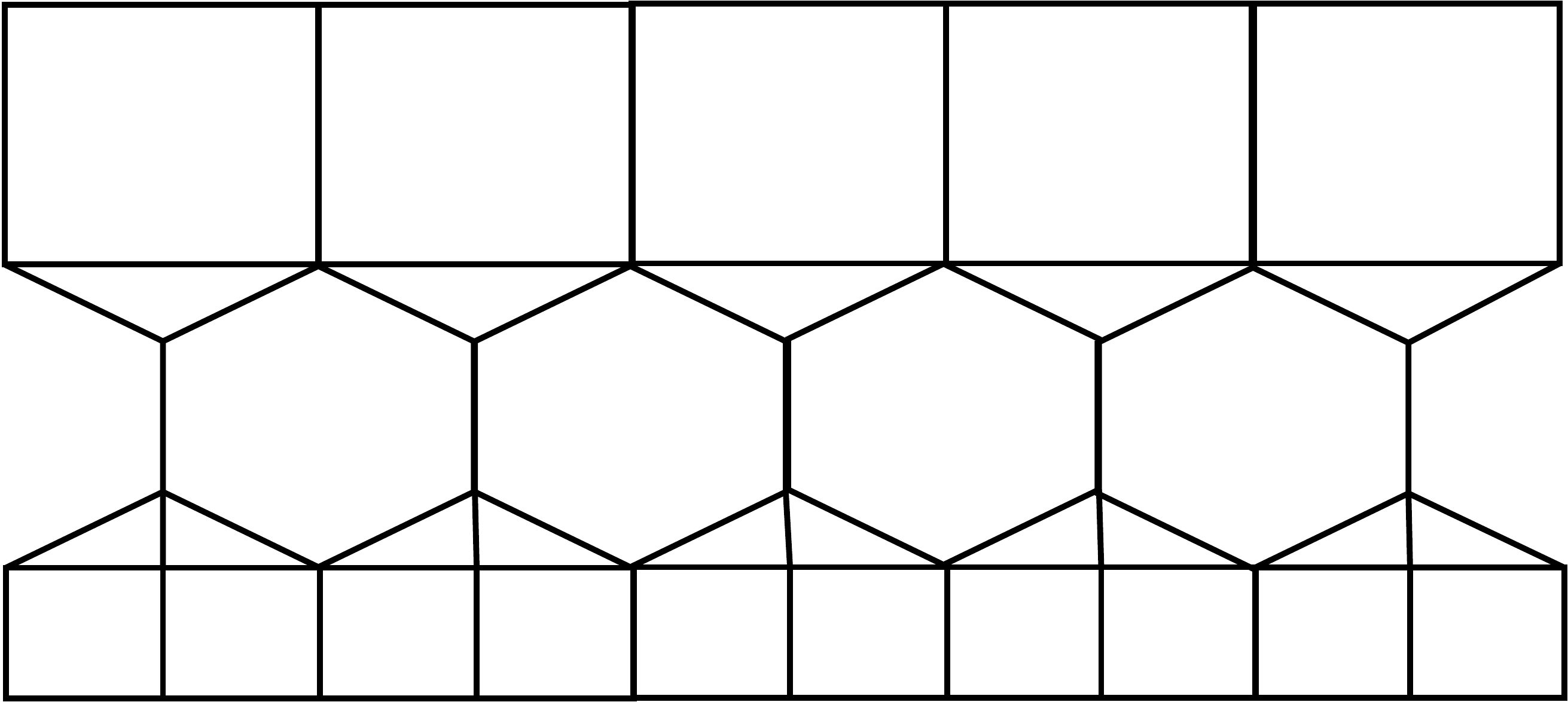}}\hspace{5pt}
\caption{Formation of polyhedral elements at hanging nodes in Cartesian mesh } \label{meshele}
\end{figure}

Furthermore, the traditional mesh methods account to have a structured grid whenever it's possible to capture the flow near bluff bodies. Also, it is essential for the mesh to be body fitted since it is vital to capture the flow patterns near the building wall accurately. Due to this, it is important to create a grid of varying sizes such that it is much finer close to the buildings.  The lately introduced “Cutcell” assembly meshing algorithm in ANSYS meshing became certainly an excellent candidate which satisfies most of these requirements. Unlike using a patch conforming tetrahedral mesh arrangement close to the building due to the capability of FLUENT code to hold both structured and unstructured grids in its solver as in \citep{huang} a cartesian Cutcell method provides a way to use a structured mesh wherever possible. This not only decreases the computational time by reducing the number of nodes thereby reducing the degrees of freedom, but also takes part in h-refinement as polyhedral elements are formed at hanging nodes between structured elements of different sizes \citep{bala} as shown in Figure \ref{meshele}.

\begin{figure}[H]
\centering
\subfloat[Unzoomed view]{%
\resizebox*{14cm}{!}{\includegraphics{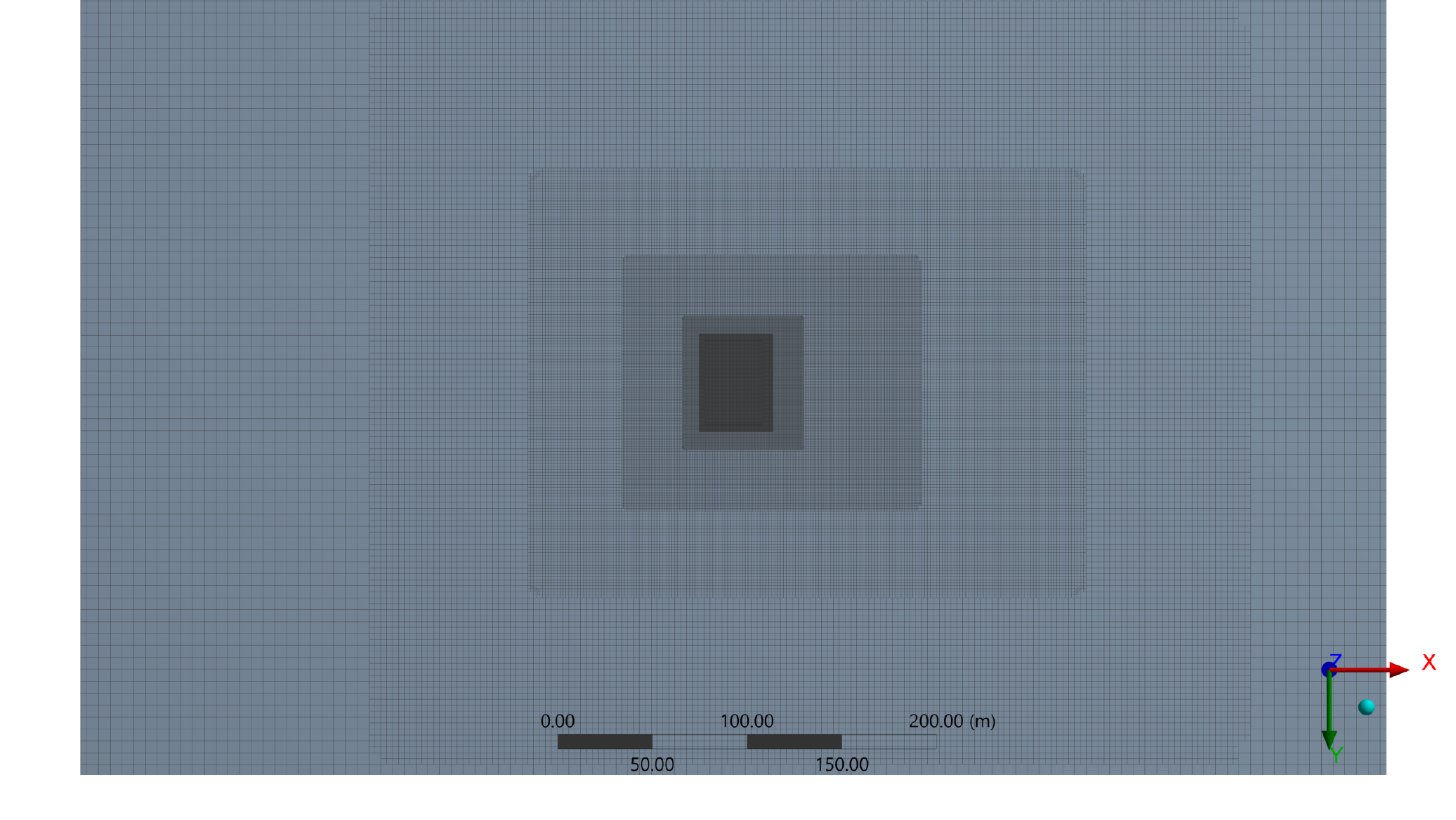}}}\\
\subfloat[Zoomed view]{%
\resizebox*{13cm}{!}{\includegraphics{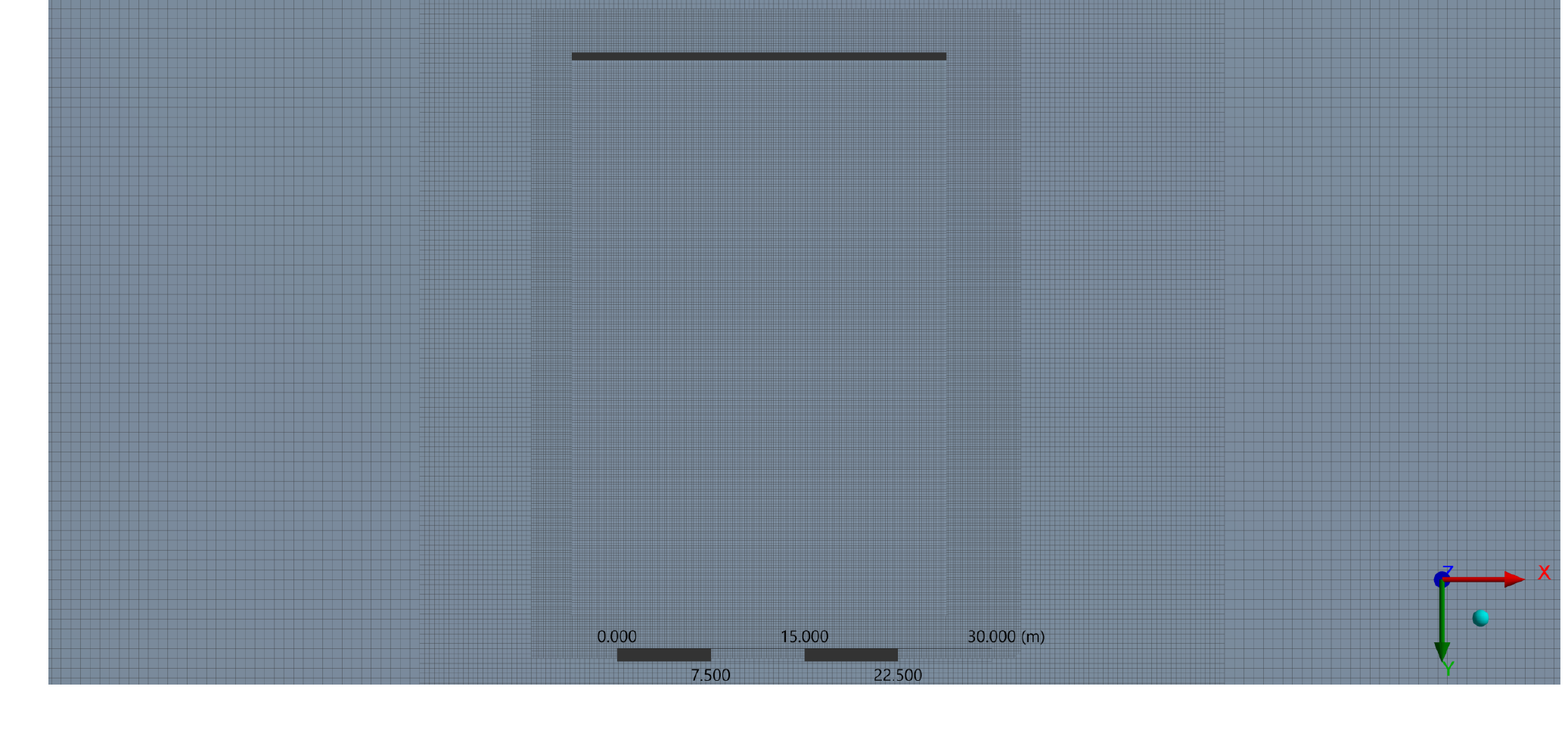}}}
\caption{Dimensions of the fluid domain} \label{mesh}
\end{figure}

\begin{figure}[H]
\centering
{\includegraphics[scale=0.3]{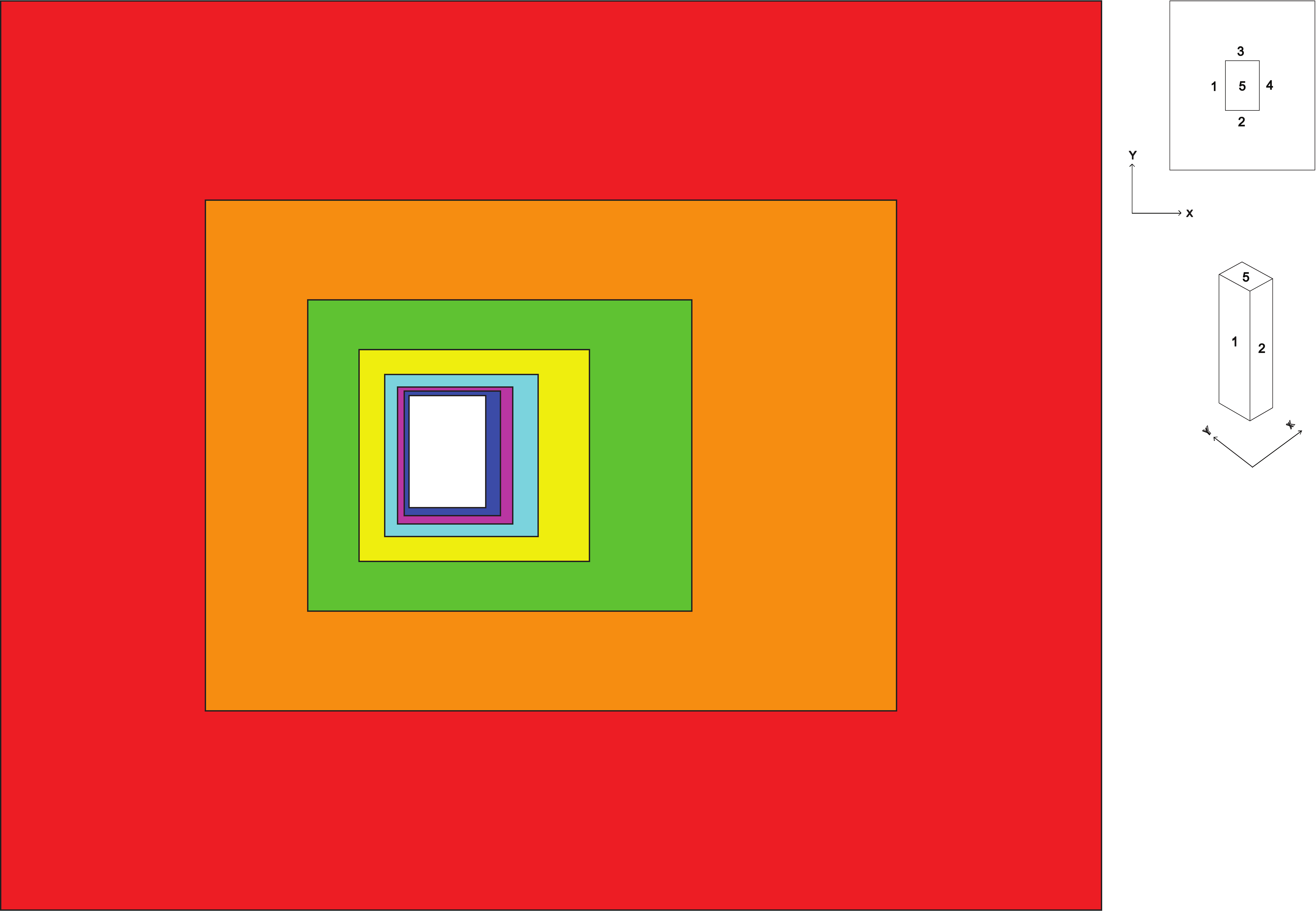}}\hspace{0pt}
\caption{Complete plan view illustration of mesh blocks} \label{meshblock}
\end{figure}

As it is seen in the mesh arrangement in Figure \ref{mesh}, a smooth transition is obtained between mesh domains of varying sizes with the combination of triangular and polyhedral elements by using the “body of influence” option available in ANSYS meshing. This also enables to keep the grid sizes smaller close to the building and paves the way to use large mesh sizes far away from the building. On the other hand, the use of such a method gives advantages over multiple body meshing since a contact meshing is formed within contact regions if traditional methods of assembly meshing are used. This would have not only increased the node count but would have displayed discontinuing streamlines due to the elements formed at the interface with negligible thickness. Table 1 gives the element size in each mesh region and Figure \ref{meshblock} shows the division of these mesh blocks.\par
To decide on the grid sizes, as described earlier the percentage of the kinetic energy-resolved by the mesh should be considered. As this study is mainly focused on finding the pressure distribution around the building accurately, the size was decided in such a way to solve the maximum amount of turbulent kinetic energy near the mesh regions very close to the building cavity. To calculate the lowest mesh size, initially, a parameter called the integral length scale needs to be calculated as shown in Equation \ref{eqn11}.

\begin{equation} \label{eqn11}
l_o =\frac{ k^{3/2}}{\epsilon}
\end{equation}

In this expression, turbulent kinetic energy solved by the mesh is denoted by \emph{k} and the turbulent dissipation rate is denoted by \textepsilon. To hold at least 80 percent of the turbulent kinetic energy, the minimum grid length should be less than at least ${1/5}^{th}$ of the integral length scale $l_0$. \par
The value of $l_0$ computed based on the turbulent kinetic energy and the turbulence dissipation rate yielded a value of 3.72 m which gives a minimum grid length of 0.74 m. However, the two blocks closest to the building cavity are of mesh sizes 0.25 m and 0.5 m respectively which means that most of the turbulent kinetic energy close to the building cavity is solved by the mesh used for the Large-eddy simulation. 

\begin{table}[H]
\tbl{Positions and the grid sizes of the mesh blocks}
{\begin{tabular}{ccccc} \toprule
& & \multicolumn{3}{l}{Distance from the building to the mesh block surface (m)} \\ \cmidrule{3-5}
Mesh block & Mesh size (m)& From surface 1,2,3 & From surface 4 & From surface 5  \\ \midrule
\cellcolor{blue!90}1 & 0.25& 2.95 & 5 & 2.5 \\
\cellcolor{magenta!90}2 & 0.4& 2.5 & 5 & 2.5 \\
\cellcolor{cyan!90}3&0.5&5&10&5\\
\cellcolor{yellow!90}4&1&10&20&10\\
\cellcolor{green!90}5&1.2&20&40&20\\
\cellcolor{orange!90}6&1.6&40&80&40\\
\cellcolor{red!90}7&3.2&80&80&60\\ \bottomrule
\end{tabular}}
\end{table}

Also since wake formation is observed behind the building, the grid was created in a way that finer mesh sizes dominate the rear side of the building than at the front as shown in block sizes in Figure \ref{meshblock}. Note that in addition to the blocks shown, the global mesh size for the remaining domain surrounding these blocks was set to 6.4 m.\par
The complete grid generated for the fluid domain consisted of 38 million elements and the computations were performed in a workstation having 24 cores and 128 GB of memory.

\subsection{Boundary conditions}
To draw out an effective comparison between experimental and simulation results, it is mandatory to obtain a clear idea about the nature of the boundary conditions used when conducting the experiment. Mainly, the inlet boundary conditions should correlate to the wind conditions used in the wind tunnel experiment to get maximum efficacy of the simulation.\par
\begin{figure}[H]
\centering
{\includegraphics[scale=0.5]{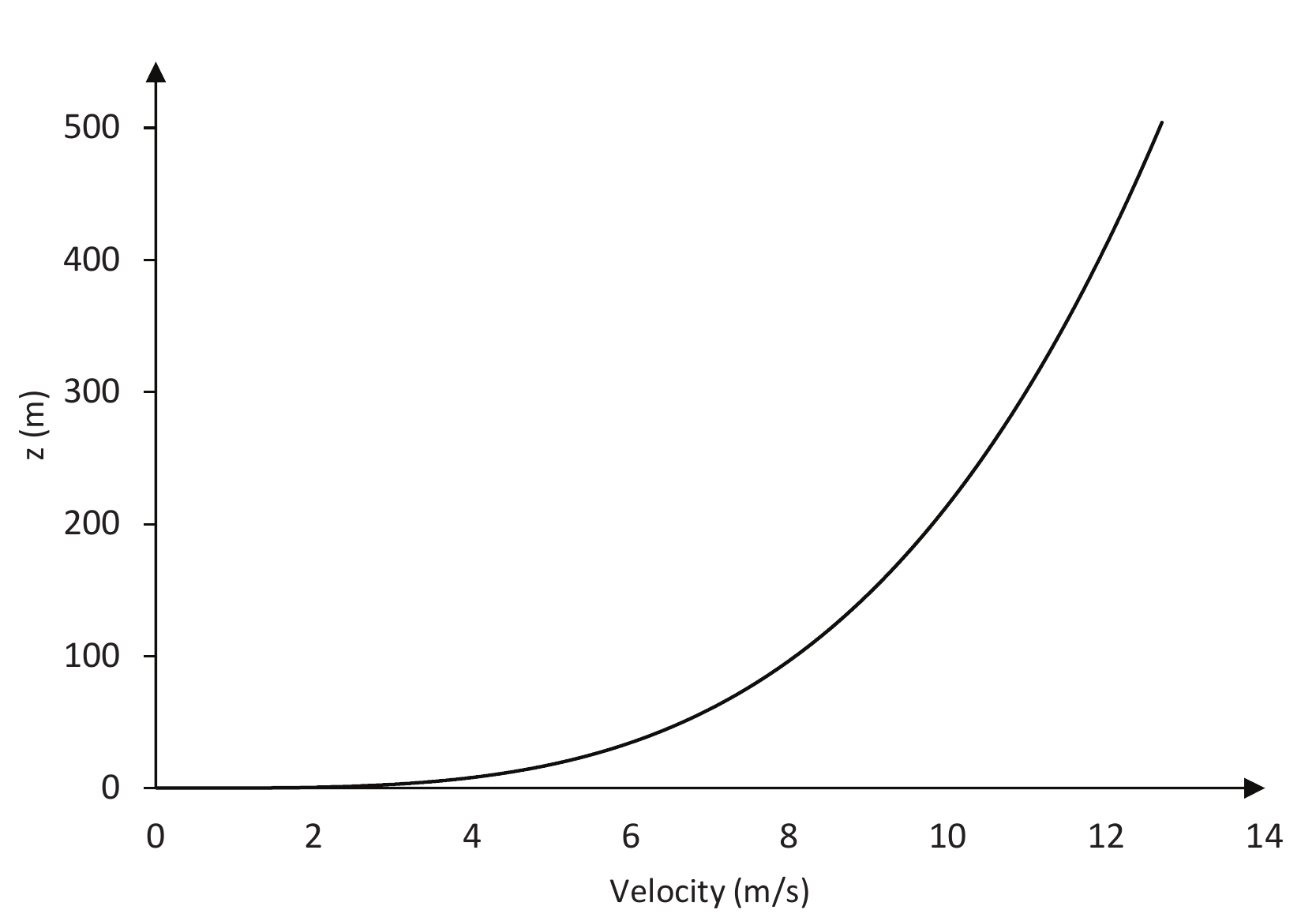}}\hspace{0pt}
\caption{ Power-law exponent profile used at the velocity-inlet} \label{veloprof}
\end{figure}

Usually, two types of inlet profiles are used for boundary layer conditions. One uses a power law exponent, whereas the other uses a function related to logarithmic laws. Different studies have used these two at multiple occasions depending on the preferences of the authors involved. In the study of \cite{oba}, a logarithmic law has been used to estimate the boundary layer, and \cite{huang} has used the power law exponent profile. In this study, the power law exponent profile has been followed according to Equation \ref{eqn12} due to the excellent comparison obtained in \cite{huang} relative to experimental readings of \cite{mel}. and the corresponding plot for the profile is shown in Figure \ref{veloprof}.

\begin{equation} \label{eqn12}
\frac{U}{U_H} =(\frac{Z}{Z_H})^a
\end{equation}

The terminology used for this expression can be summarized as follows. \par
\emph{U} gives the wind velocity at the preferred height; $U_H$ gives the speed of the wind at reference height. (maximum height of the domain); \emph{Z} denotes the preferred height and $Z_H$ denotes the maximum height of the domain. Finally ‘\emph{a}’ gives the power law exponent which has been estimated to be 0.28 in \cite{mel}.

\subsection{Computational domain and mesh arrangement for structural domain}

The significance of this study is mainly to determine the effect generated by the Aeroelastic nature of the building. The building structure was modelled as a simple hollow cantilever beam of rectangular shell cross section fixed to the ground. The thickness of the shell was determined such that the building shows a fundamental natural frequency of 0.2 Hz according to \cite{huang}. \par
The shell structure was modelled using cladding as the material of Modulus of Elasticity of 25 GPa. To further emulate the behavior of a building, a combination of different mesh elements was used for the building structure. The shell was modeled using standard shell 181 elements due to the advantages it gives with respect to solution time, operational capacity, stability and accuracy in contrast to the use of solid elements \citep{irsel}. Moreover, 39 interior floors touching the four walls were created to enhance the structural stability, which were also modeled using 4 node 6-DOF shell 181 elements. Furthermore, columns were embedded in addition to the plates and the building shell.  The resulting structure showed rigidity to forces of the magnitude of wind loads and the non-linear deformations which were present earlier were eliminated. The dimensions of the building components were optimized such that the concrete plates have a thickness of 0.85 m leading to a fundamental frequency of 0.2 Hz obtained for the vibration. .

Consequently, it was observed that the mode shape for the fundamental frequency of 0.2 Hz corresponded to the direction in which the deformation acts once the wind load acts on the building surface and the modal frequencies are given in Table 2.

\begin{table}[H]
\tbl{Frequency modes}
{\begin{tabular}{ccc} \toprule
Mode&&\hspace{0.5cm}Frequency(Hz)  \\ \midrule
1&&\hspace{0.4cm} 0.20 \\
2&&\hspace{0.4cm} 0.26 \\
3&&\hspace{0.5cm}0.92\\
4&&\hspace{0.5cm}1.05\\
5&&\hspace{0.5cm}1.10\\
6&&\hspace{0.5cm}1.17\\ \bottomrule
\end{tabular}}
\end{table}

\section{Two-Way Fluid Structure Interaction}
\subsection{Overview}

As far as two way fluid structure analysis is concerned, there could be two approaches; either a monolithic approach or a partitioned approach \citep{akbay} which is used more commonly. The ANSYS Workbench uses a partitioned coupling framework that enables support for data mapping in shared memory and distributed parallel processors which makes simulation times lower \citep{chima}. \par

A brief description of the theoretical background relevant to this study is provided in this section for ease of understanding. For more information it is highly recommended to refer \cite{anssc}.
Two algorithms are usually used for the data transfer between two different setups.
\begin{enumerate}
\item Profile preserving algorithm - Used to transfer non-conserved quantities (eg- displacement)
\item Conservative profile preserving algorithm - Used to transfer conserved quantities (eg- force) 
\end{enumerate}

Both these algorithms work under four sub algorithms.
\subsubsection{Data Pre-processing}
These algorithms are used in the initial stage to create supplemental data on mesh locations which are required at mapping and interpolation stages.
\subsubsection{Mapping} 
Mapping algorithms are also divided into two based on the type of data transfer algorithm. Thus there exist Bucket surface and General Grid Interface (GGI) algorithms that are dedicated to Profile preserving and Conservative algorithms respectively. \par
The Bucket surface algorithm is employed to assign weights required for displacement mapping in this study. Here the mapping takes place between the nodes on the target side with mesh elements on the source side. Therefore the initial step is to focus on dividing the mapping source mesh into an imaginary structured grid; each grid section here is called a “Bucket”. As the system coupling algorithm does not support interactions with 2D buckets, bucket grids are generated only for a 3D grid. Subsequently, every node would be linked with a generated bucket. To ensure proper linking takes place, a checking loop is activated which traces all source elements to find target elements. This isoparametric mapping takes place according to the nonlinear Equation \ref{eqn13} given below.

\begin{equation} \label{eqn13}
{\hat{x}}=[N_a^e(\zeta)]\{x_a^e\}
\end{equation}

In this expression, \textzeta\ acts as vector element local-coordinates which correspond to global coordinates of the target node which is $\hat{x}$. Also, $N_a^e$ denotes the linear shape functions corresponding to the source element, whereas $x_a^e$ denotes global coordinates of element local node $a$. This mapping algorithm also includes tolerance levels when mapping between nodes and elements and more details about these can be found on \cite{jansen}.\par
The General Grid Interface algorithm is used for force mapping functions in a way such that element faces on both source and target sides are divided by the number of nodes, and the resulting faces are converted into 2D quadrilaterals which are eventually turned into “control surfaces”. The final weight contributions for mapping are evaluated through an accumulation of these control surface contributions. More details on this algorithm can be found on \cite{galpin}.

\subsubsection{Interpolation}

This is where target node values are provided with the aid of source data and mapping weights are generated by the use of mapping algorithms. As given in Equation \ref{eqn14} mapping weights are applied.

\begin{equation} \label{eqn14}
\phi=\sum_{i=1}^n w_i \phi_i
\end{equation}

$\phi$ denotes the target node, thereby making $\phi_i$ the value at $i$ th source node. $w_i$ denoted the associated weight. Value of $n$ depends on the mapping algorithm used. Accordingly, for the bucket surface algorithm, $n$ implies the source element node number while for GGI mapping it denotes the number of intersected areas on the control surface.\par
After this stage, Post processing algorithms are used on interpolated data to increase the convergence. Two algorithms are usually used for this purpose which could be a ramping or under-relaxation algorithm; both of which are explained in \cite{anssc}. Collection of all these algorithms would complete the data transfer algorithm. After each iteration undergoes all the steps in the data transfer algorithm, the root mean square is checked for data transfers in both ways. This RMS value is calculated according to Equation \ref{eqn15} .

\begin{equation} \label{eqn15}
RMS=\sqrt{\overline{(\hat{\Delta_l}})^2}
\end{equation}

In here, the parameter $\hat{\Delta_l}$ is the normalized data transfer value which is itself calculated as in Equation \ref{eqn16}.

\begin{equation} \label{eqn16}
\hat{\Delta_l}=\frac{\Delta_l}{0.5\times((max|{\phi}|-min|{\phi}|)+\overline{|{\phi}|})}
\end{equation}

Where $\phi$ and $l$ denotes the values and locations of the data transfer values respectively.
This is the general data transfer code that is used in the system- coupling database which is employed in this study. It should be noted that these algorithms could be used only for triangular and quadrilateral elements between two data exchanging surfaces. Although polyhedral elements were used for this study, none of them were included between mesh grid transitions, which led to the successful compatibility of the algorithms to the grid generated.

\subsection{Functioning of two way- FSI}

Initially, individual solvers relevant to Structural and fluid setups are modified to enable the coupling between the two domains. Thus, the building model in the Structural solver is arranged in a way such that the building floor is fixed to the ground and the building walls act as fluid system couplings. Five couplings are formulated for the five walls and once the simulation is initiated these walls transfer data with the fluid domain according to Bucket surface and GGI algorithms . Moreover, in the fluid domain setup, the entire mesh is set to a deformable dynamic mesh so that the grid has the capability to keep contact with the fluid domain at each time step involving structural deformation. Similar to the structural setup, the fluid domain building walls are also set to “system coupling” regions so that the solver identifies with what domains it needs to transfer data in the structural setup. \par
\begin{figure}[H]
\centering
{\includegraphics[scale=0.3]{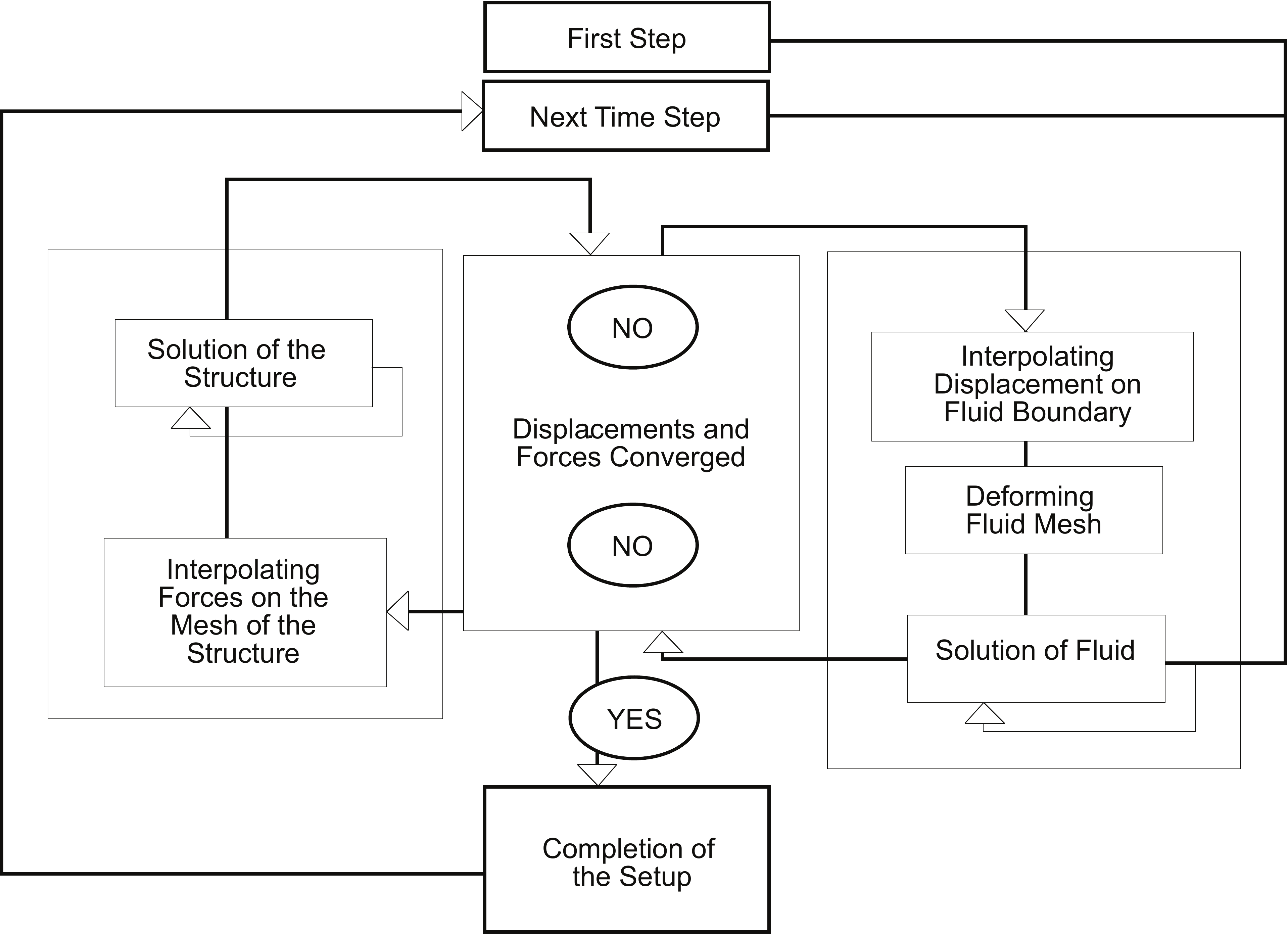}}\hspace{0pt}
\caption{ Flow diagram for the iterative scheme in Ansys system coupling} \label{fig10}
\end{figure}

Both these setups are centrally governed by the system coupling system and so that the individual solvers of fluid analysis and structural analysis are shared within this setup. Here, mainly data transfers are created such that the forces on the fluid section are transferred towards the structural domain according to GGI algorithm while in the same iteration displacement data from the structural setup are shared with the fluid domain according to Bucket surface algorithm. Thus this process repeats itself for each timestep until every data transfer is converged leading this to a fairly time-consuming process when compared to the rigid models. This iterative process is described by the flow diagram in Figure \ref{fig10}. \par

In addition to these data a coupled pressure solver and a second order transient formulation were used for the solving process of the equations involved. A time step of 0.1 s was able to achieve convergence for all time steps and the setup was allowed to run for 150 time steps to get a stabilised pressure variation along the building wall. 

\section{Results and Discussion}

The results of the three cases :  Rigid RANS model analysis, Rigid LES model analysis and aeroelastic LES model analysis were taken adhering to the main objective; which is to make a comparison between simulation models to find out which gives the closest approximation to the experimental phenomena such that costly experiments can be replaced by simulations in future at least for the preliminary level of design. For this purpose, the results were mainly compared with experimental results obtained from the \cite{mel} study at 2/3\textsuperscript{rd} the height of the building and from \cite{goli} study at the 1/3\textsuperscript{rd} height. A comparison is initially drawn with respect to the normalized pressure variation (Equation 17) obtained around the building which would in turn capture all three responses related to a study on wind loads.

\begin{equation} \label{eqn17}
Normalized\,Pressure (P')= \frac{p}{\frac{1}{2}\rho u^2_H}
\end{equation}

Where $P$ is the dynamic pressure observed around the building, $\rho$ is the density of air and $u_H$ gives the velocity of the wind boundary layer at the top of the building. 

\subsection{Case 1 – Rigid RANS model analysis}

\begin{figure}[H]
\centering
{\includegraphics[scale=0.80]{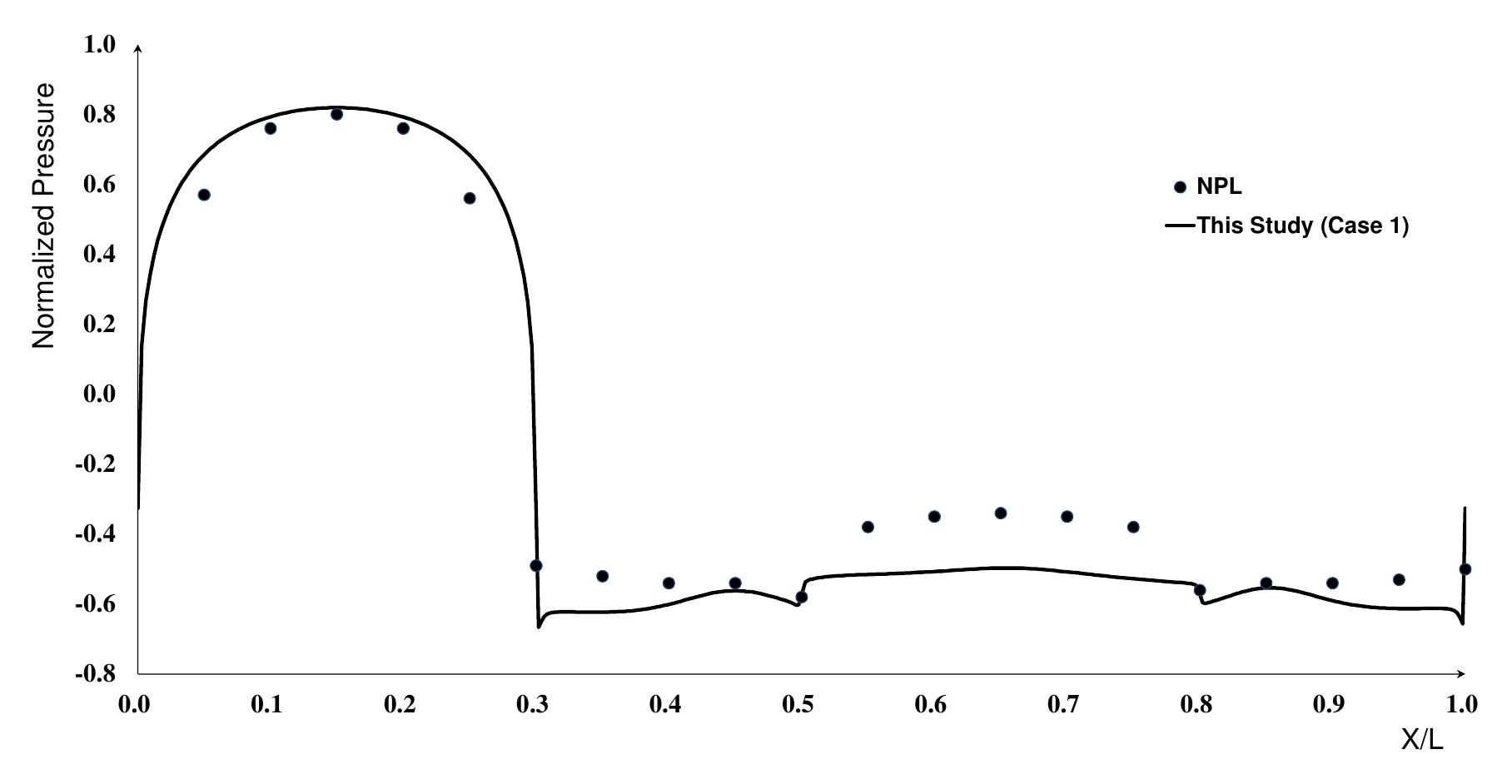}}\hspace{5pt}
\caption{Normalised pressure variation along the building wall at a height of 120 m for Case 1} \label{rans}
\end{figure}

Results obtained for the normalized pressure coefficients at 2/3\textsuperscript{rd} of the height of the building were plotted against \emph{x}/\emph{L} as shown in Figure \ref{rans} and compared them with results obtained from the NPL study of \cite{mel}. Only the NPL study has been used from \cite{mel} since it was done using the closest boundary conditions to the present study and it has shown the closest resemblance to simulation results in previous studies.\citep{huang,braun}

\begin{figure}[H]
\centering
{\includegraphics[scale=0.8]{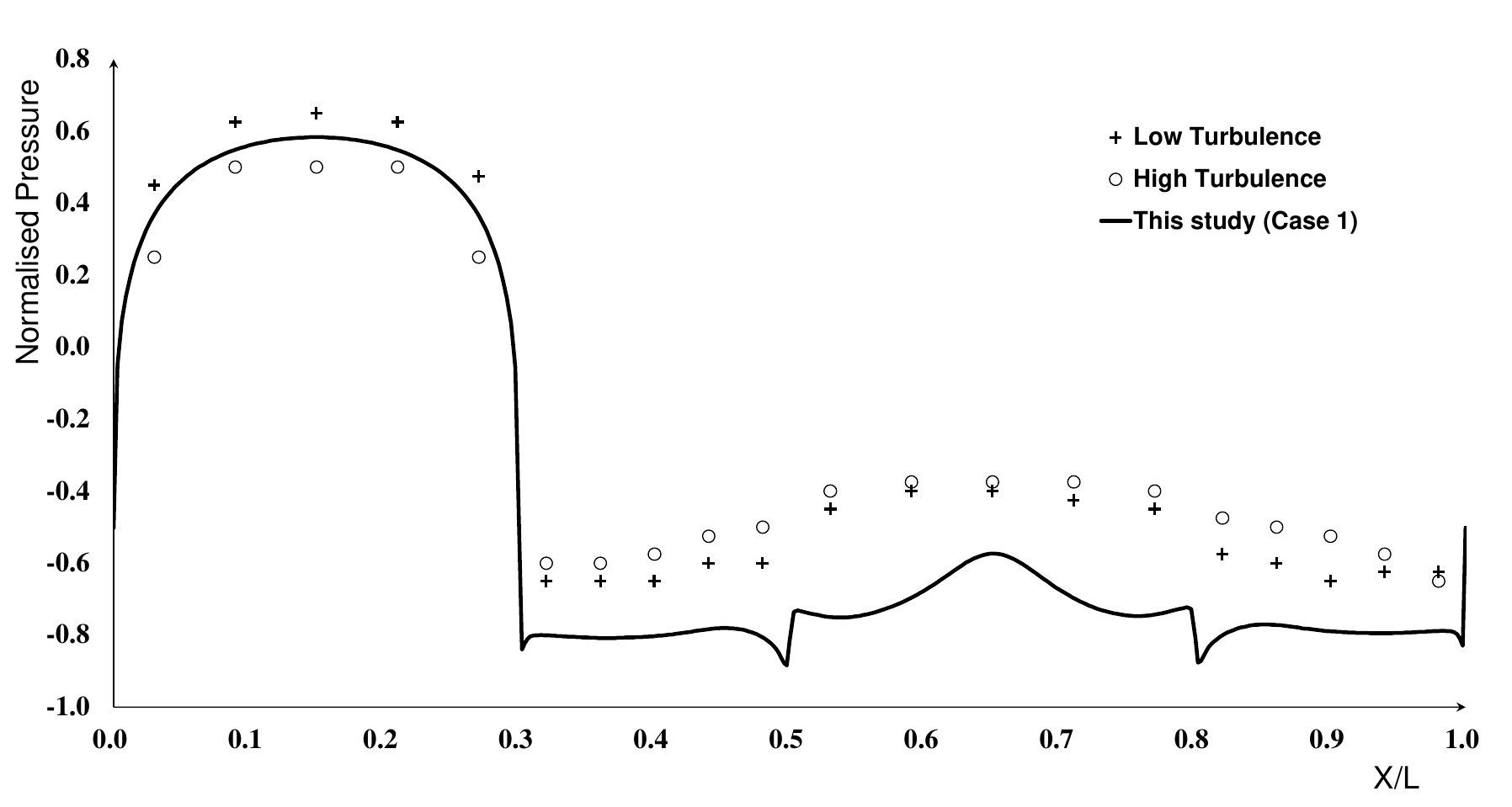}}\hspace{5pt}
\caption{Normalized pressure variation along the building wall at a height of 60 m for Case 1} \label{rans13}
\end{figure}

Overall, it is clear that along most of the faces the normalized pressure has been overpredicted to the results obtained from the \cite{mel} study. In total, along the 4 faces, about 4 experimental readings from the \cite{mel} study have been superimposed on the simulation curve obtained for the RANS study. Although the results have been predicted very precisely in the front face of the building, the rear side of the building has been unable to exhibit accurate results. There is a substantial overprediction of. This is probably because of the inability of RANS to correctly predict the wake formation behind the building. This is most often the case of the RANS models that happen due to eliminating fluctuations in turbulence due to averagings. This can be more clearly seen by comparing the results at another building height (60 m from the ground) as in \cite{goli}. This study has fewer experimental points and was done under two turbulent intensities and the comparison with these results at 1/3\textsuperscript{rd} the height of the building is shown in Figure \ref{rans13}.

From Figure \ref{rans13}, it is clear that the \emph{k}-\textepsilon\ model has the ability to predict the pressure variation at the front face at multiple building heights. However, within the vicinity of other faces the deviations from the experimental results have been significant despite the correlation observed in the shape of the curve. To obtain the reasoning behind this, further investigation was carried out by looking at the eddy formation and turbulence close to the building.
\par
\begin{figure}[H]
\centering
\subfloat[Sectional front view]{%
\resizebox*{12cm}{!}{\includegraphics{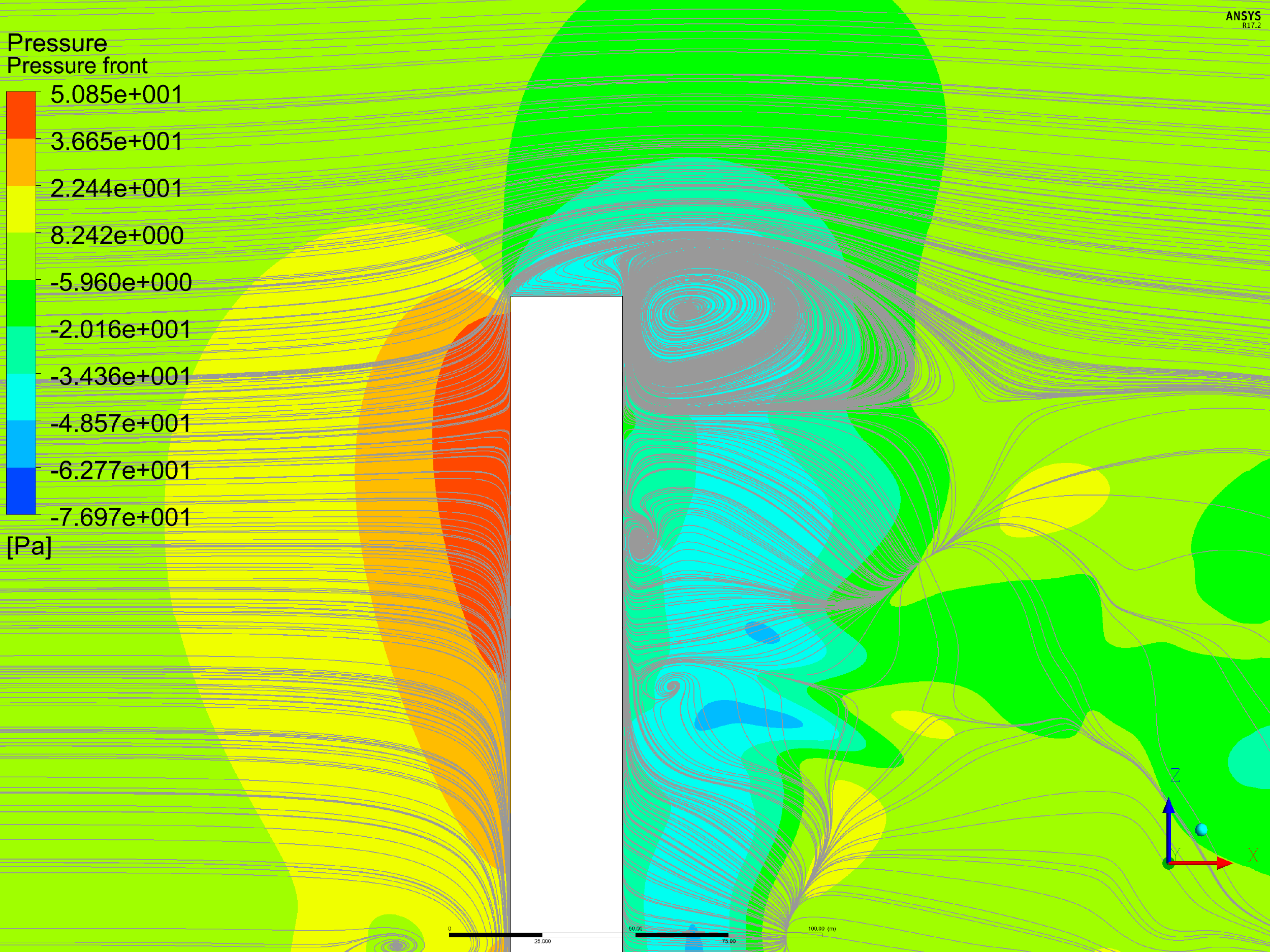}}}\hspace{5pt}
\subfloat[Sectional plan view at building height of 120 m]{%
\resizebox*{12cm}{!}{\includegraphics{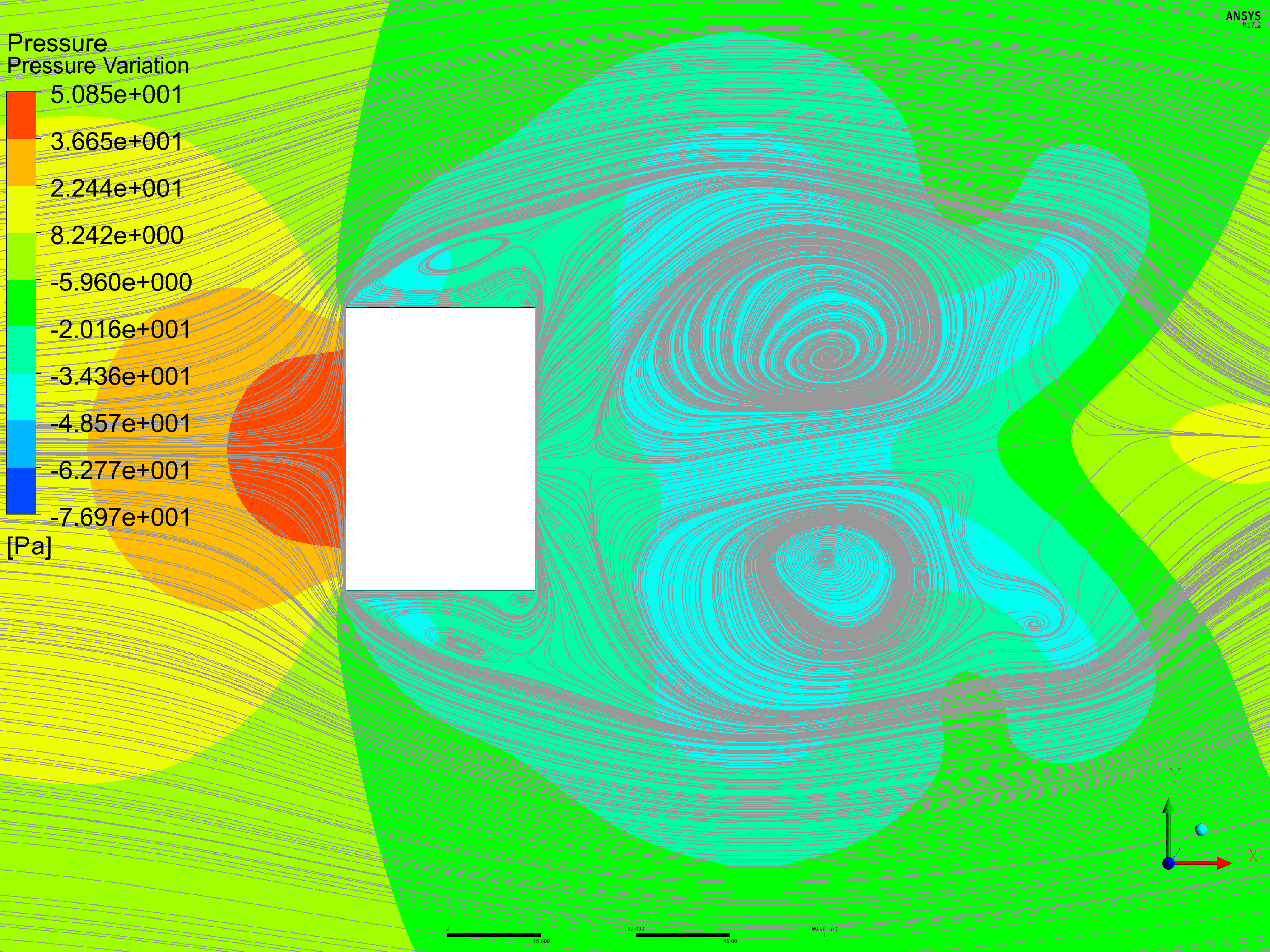}}}
\caption{Velocity streamlines and pressure variation observed for Case 1 } \label{ransview}
\end{figure}

Thus, the plot for velocity streamlines and the pressure contours at the sectional front and plan views are plotted in the same figure as shown in Figure \ref{ransview} (a) and (b), respectively.

A very high positive pressure is obtained once the wind comes and strikes the frontal face of the building thereby resulting in flow separation along building edges. This results in a positive pressure region in front of the building and also a recirculating region in the wake area which ultimately results in a pressure difference creating a net force in the along-wind direction. It is also clear that the flow separation in the frontal face happens in two directions whereas most of the streamlines flow down along the building converging into a single eddy and the flow striking very close to the top of the building tends to separate towards the wake region forming a large eddy at the top end of the building. It is further clear the streamline flow observed in the plan view is very close to being symmetric and two clear large eddies are formed. This is the main reason to observe a symmetric normalized pressure variation and hence predicts the pressure at both sides of the building to be equal, which is inaccurate in reality. However, even if the details are lesser, this simulation run took only 1/10\textsuperscript{th} of the computational effort when compared to running the solution with subgrid-scale models. 

\subsection{ Case 2 -Rigid LES model analysis}

\begin{figure}[H]
\centering
{\includegraphics[scale=0.8]{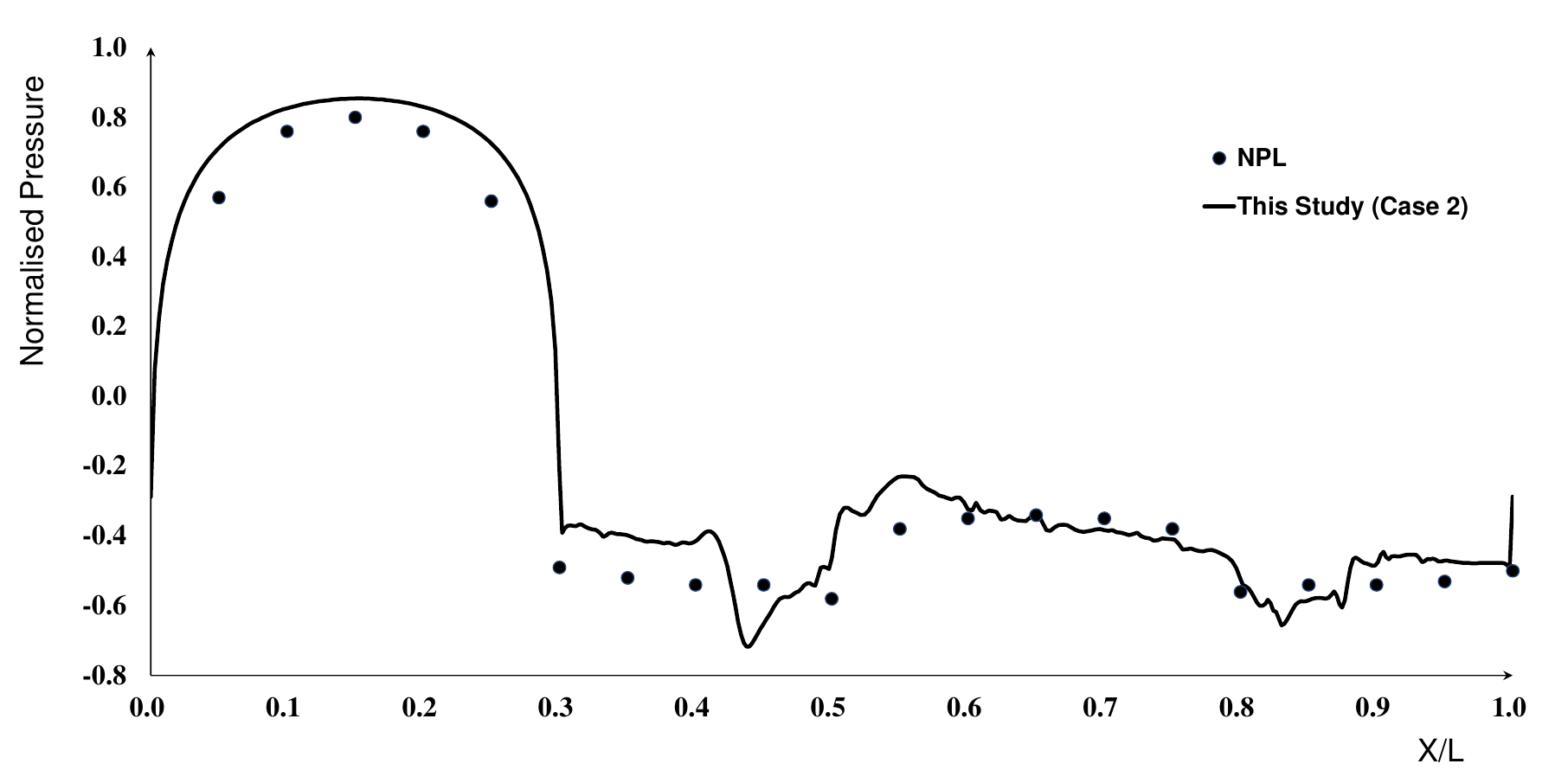}}\hspace{5pt}
\caption{Normalized pressure variation along the building wall at a height of 120 m for Case 2} \label{lesrigid}
\end{figure}

\begin{figure}[H]
\centering
{\includegraphics[scale=0.8]{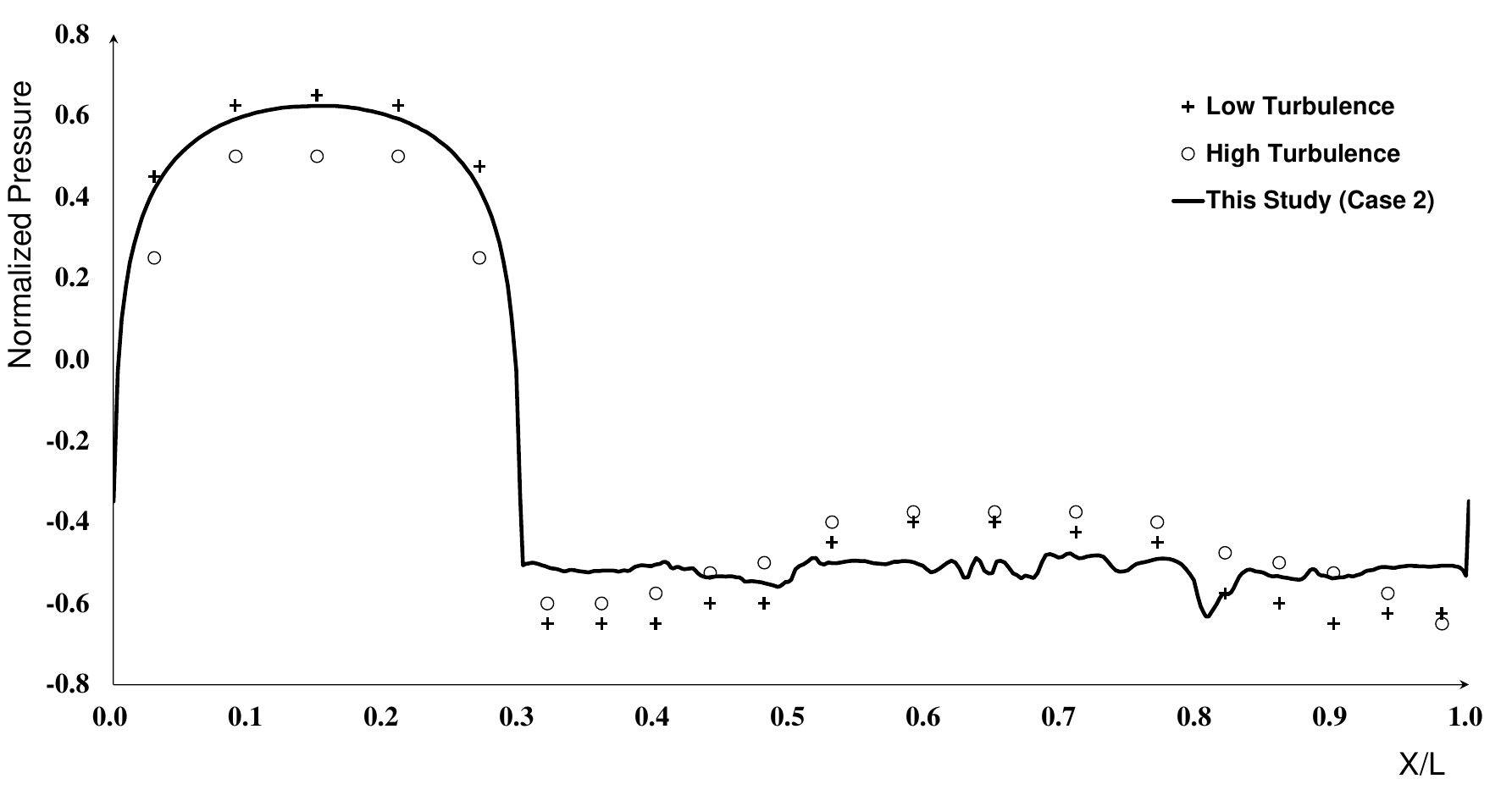}}\hspace{5pt}
\caption{Normalized pressure variation along the building wall at a height of 60 m for Case 2 } \label{lesrigid13}
\end{figure}

Figure \ref{lesrigid} and Figure \ref{lesrigid13} show the pressure variation of the building at 120 m  and 60 m height respectively once a rigid model was simulated by using the Smagorinsky SGS turbulence model. 

\begin{figure}[H]
\centering
\subfloat[Sectional front view]{%
\resizebox*{9cm}{!}{\includegraphics{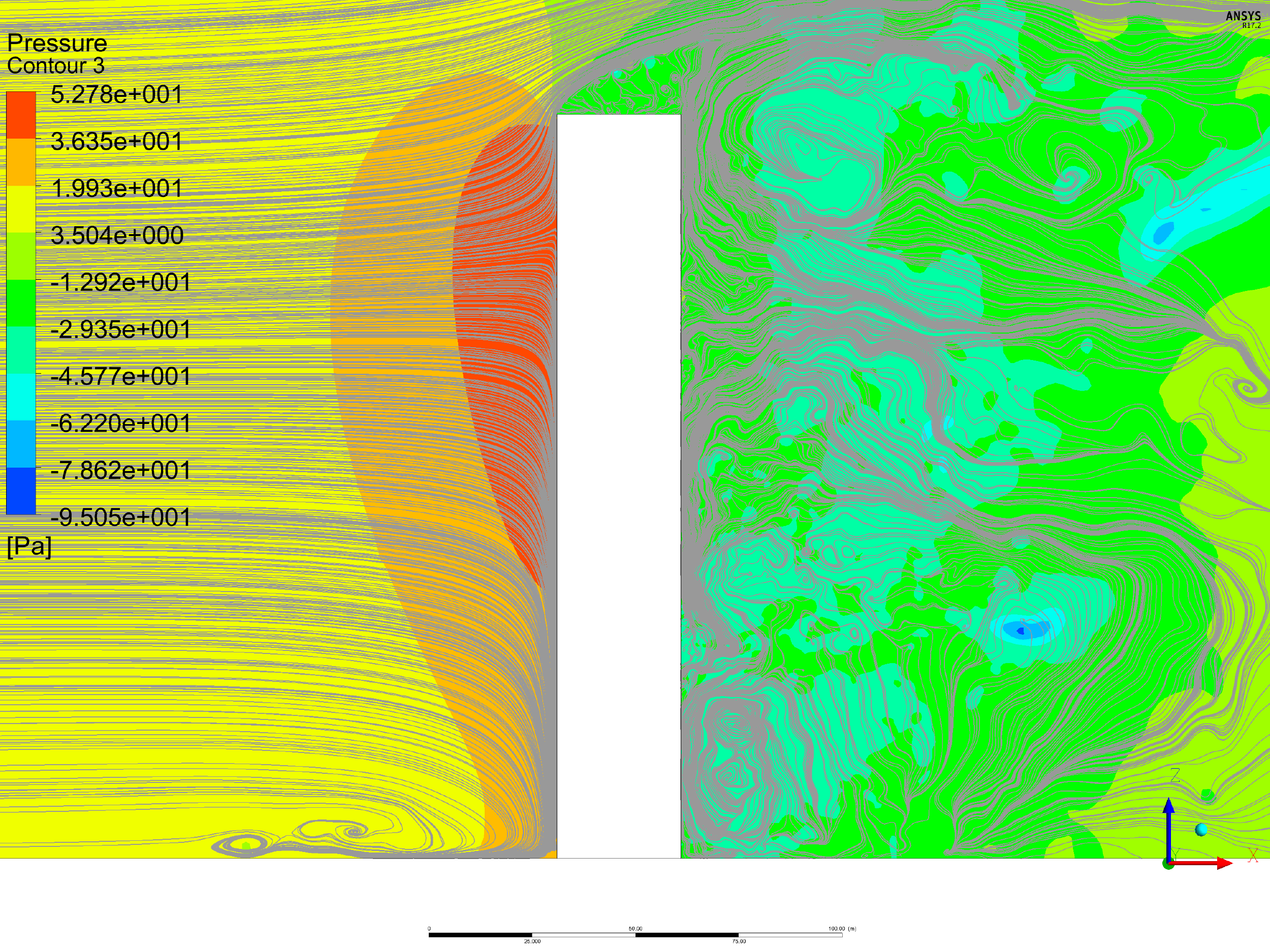}}}\hspace{5pt}
\subfloat[Zoomed sectional front view]{%
\resizebox*{9cm}{!}{\includegraphics{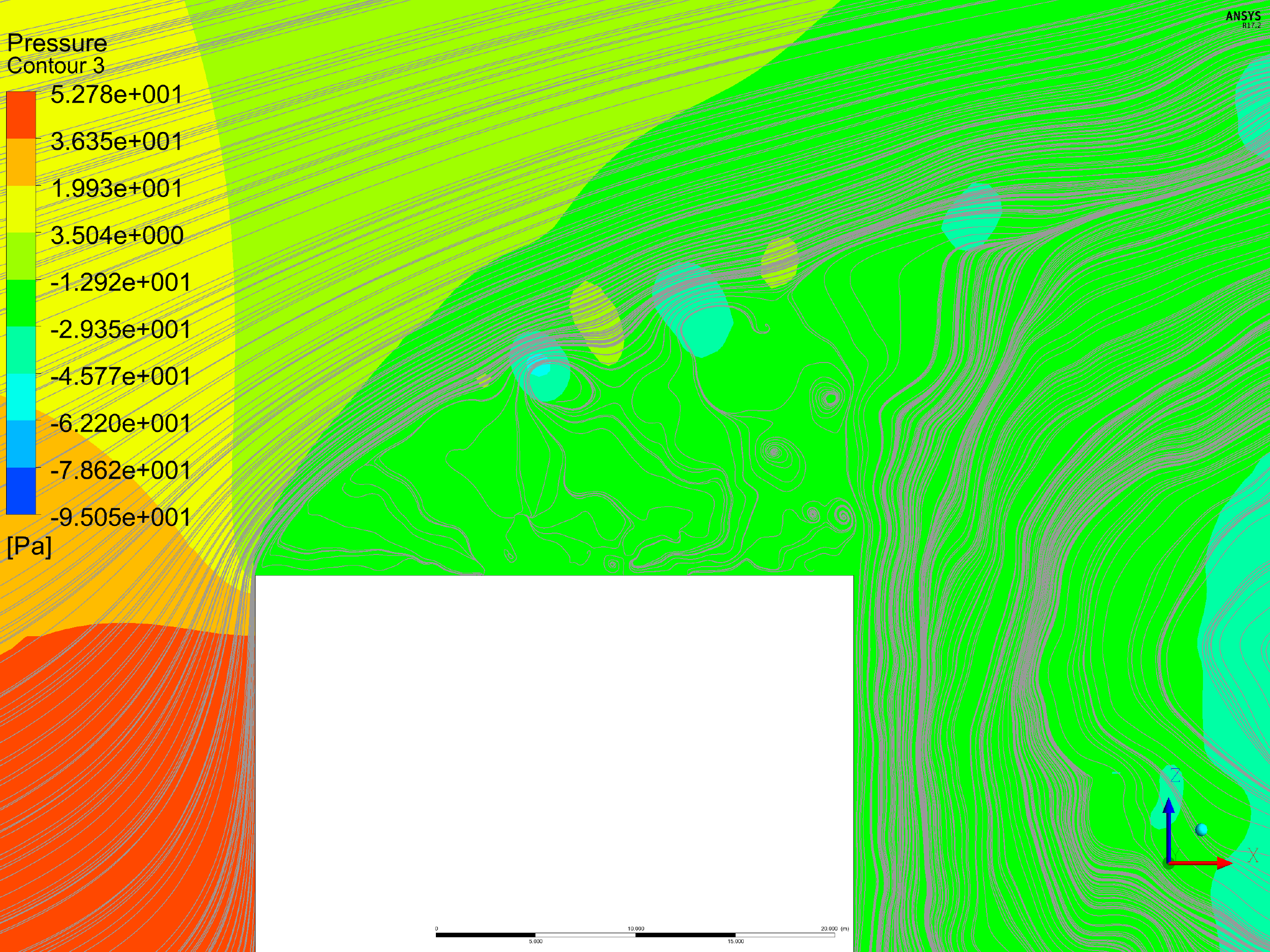}}}\hspace{5pt}
\subfloat[Sectional plan view at building height of 120 m]{%
\resizebox*{9cm}{!}{\includegraphics{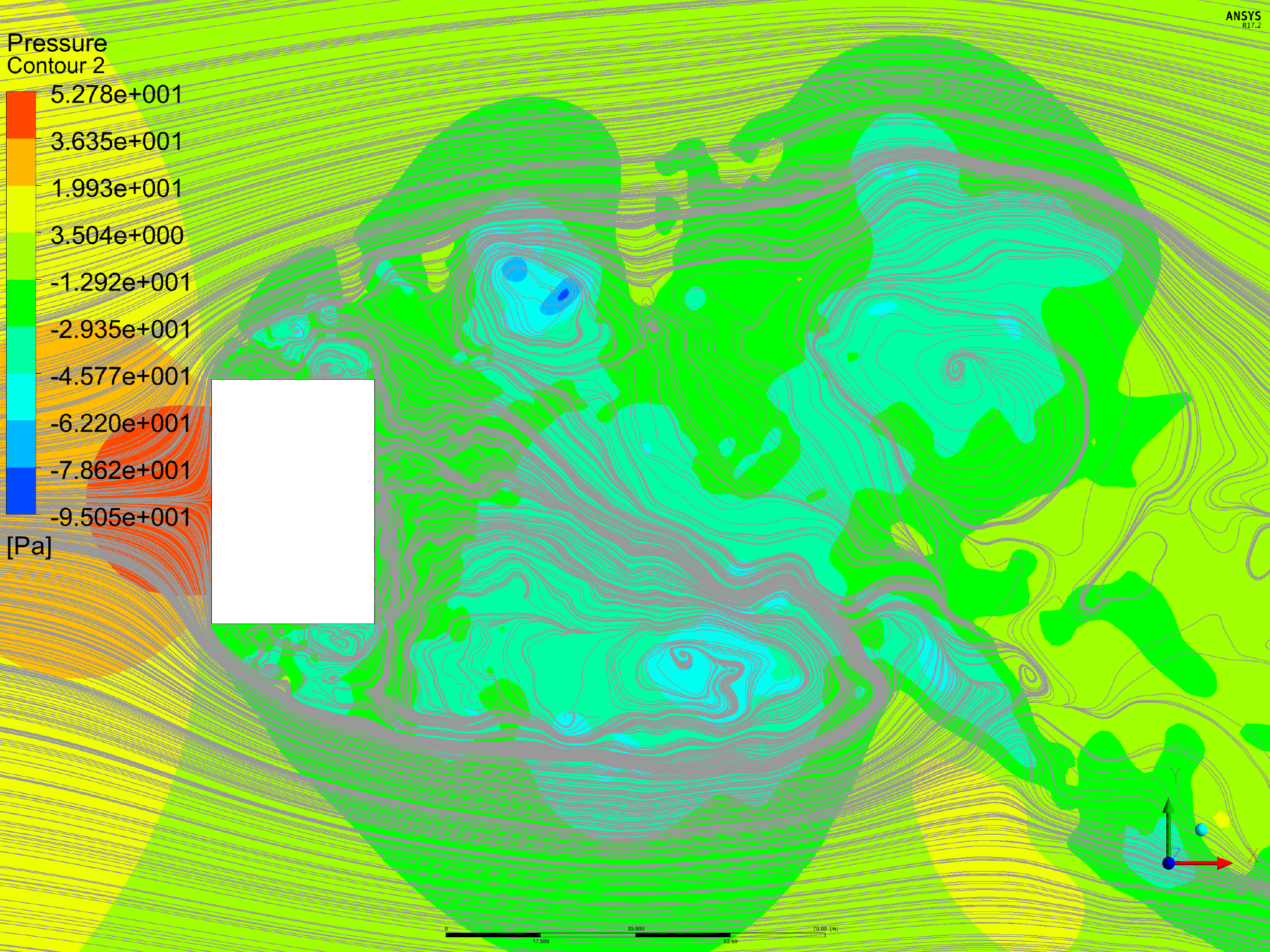}}}
\caption{Velocity streamlines and pressure variation observed for Case 1 } \label{lesrigidview}
\end{figure}

In Figure \ref{lesrigid}, roughly 8 points from the simulation data have been successfully superimposed to the experimental data for the NPL study suggesting better correlation to the RANS model.  At first glance, one would observe that as in the RANS model the front face of the building shows great alignment with all the experimental readings. Although, the dip in normalized pressure at the first side of the building has been overestimated in the LES rigid model by a fairly large margin, the pressure at the back wall has been predicted to a high accuracy level in the LES analysis suggesting that the wake formation behind the building has been captured to high precision. This is further elaborated in the velocity streamline and pressure contour formation depicted in Figure \ref{lesrigidview}. 

Similar to the Case 1, Upon striking the building wall, air streamlines separate into two directions such that eddies are formed close to the ground from the streamlines that travel down. However, the contrasting behavior is seen in flow separation at the top building edge as shown in Figure \ref{lesrigidview} (a). Upon zooming into the building top it was observed that a large number of tiny eddies are formed near wake separation suggesting a high percentage of turbulence resolvement. Moreover, due to this high amount of small amplitude vibrations could result at the top of the building which would not be captured by a rigid building structure. The plan view in Figure \ref{lesrigidview} (c) also shows an unsymmetric eddy distribution of various sizes amalgamating to form a more complete wake region of lower pressure values.

\subsection{Case 3– Aeroelastic LES model analysis}

\begin{figure}[H]
\centering
{\includegraphics[scale=0.8]{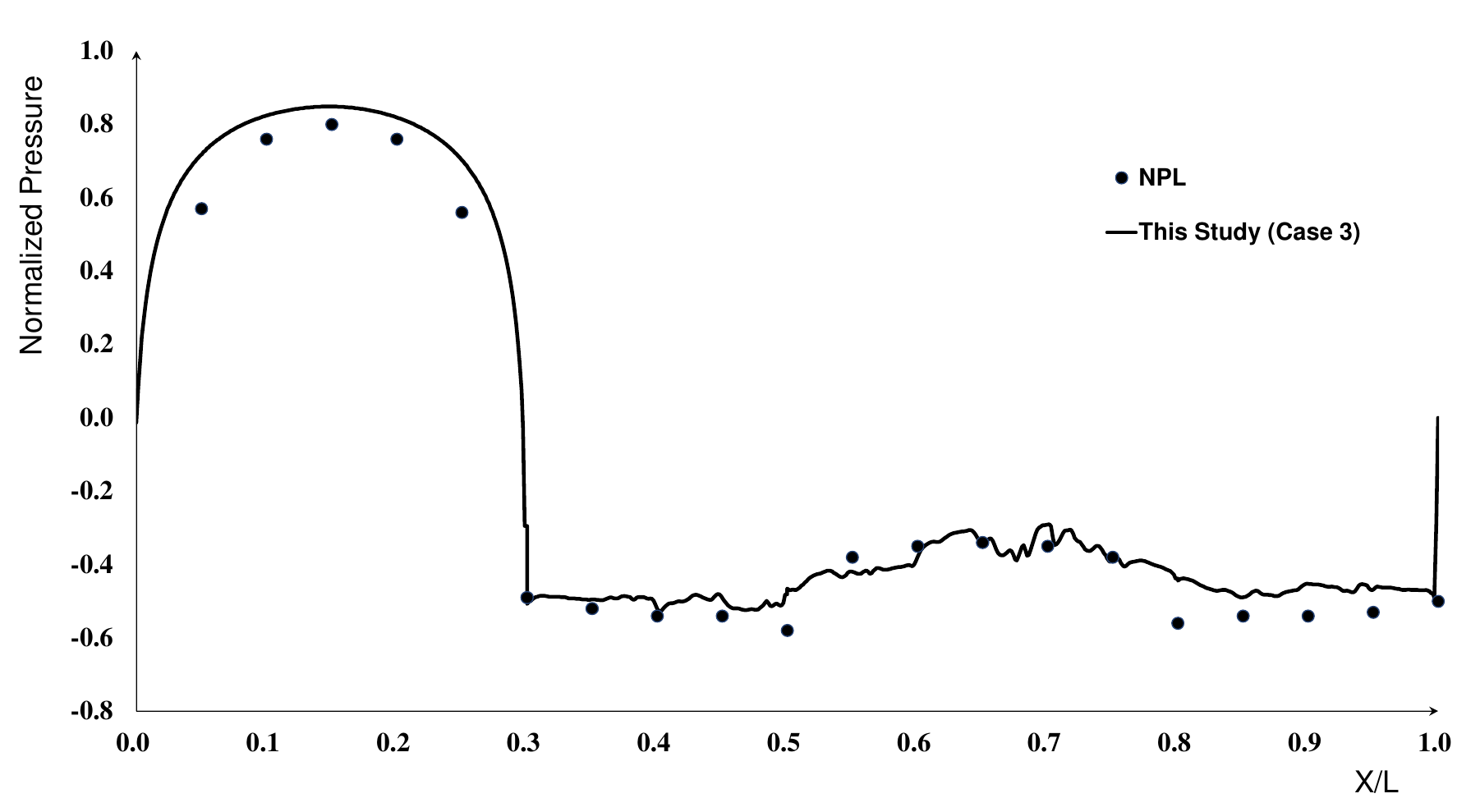}}\hspace{5pt}
\caption{Normalized pressure variation along the building wall at a height of 120 m for Case 3} \label{lescoupled}
\end{figure}

To plot the results of pressure variance when the building undergoes deformation the usual approach of plotting the pressure along the building wall line could not be used. This is because due to the aeroelastic nature the position of the building would be different from the initial position and its coordinates are unknown. Hence a plane was defined at a building height of 120 m and the pressure variation was plotted along the line (polyline) at which this boundary intersects with the building. The normalized pressure variation at 2/3\textsuperscript{rd} of building height was obtained accordingly and plotted as in Figure \ref{lescoupled}.

\begin{figure}[H]
\centering
{\includegraphics[scale=0.8]{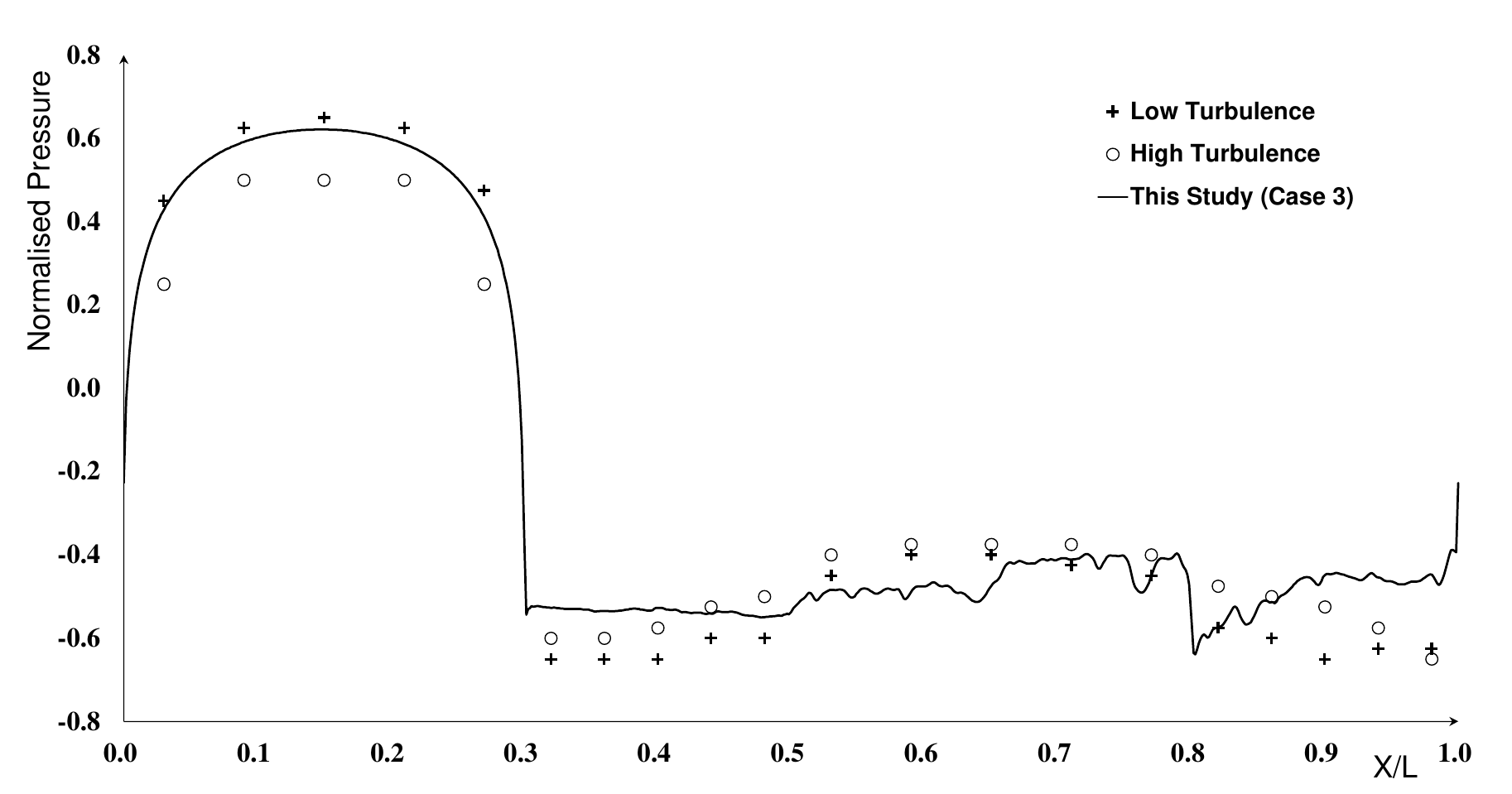}}\hspace{5pt}
\caption{Normalized pressure variation along the building wall at a height of 60 m for Case 3} \label{lescoupled13}
\end{figure}

In this variation, in Figure \ref{lescoupled}, it is clearly seen that the pressure variation along the Z=120 line has been able to trace a high majority of points for readings obtained for NPL in \cite{mel}. Thus, in total, the curve has been able to recreate 12 of 15 data points in the NPL experiments conducted which supersedes the precision shown by \cite{huang} and \cite{braun} studies. Furthermore, the study of \cite{goli} determined experimental pressure values predicted at the 1/3 \textsuperscript{rd} of the building height and it has also been used to determine the degree of accuracy of this aeroelastic study as in \cite{braun} as in Figure \ref{lescoupled13}.\par
According to Figure \ref{lescoupled13}, the pressure variation in the front face lies midway between the low turbulence and high turbulence experiments conducted by Goliger and Milford (1988). This is similar to the pressure behavior at the front face observed in \cite{braun} despite the small over-prediction in \cite{braun}. However, this may be due to the change in the considered Reynolds number for that study and the present study. The almost constant pressure approximation in the side face matches with the results obtained in \cite{braun} where the simulation curve lies close to the experimental data. However, the major difference in the two studies is within the rear wall boundary of the building. As observed in \cite{braun} the pressure variation at the rear side seem to deviate largely from the results gained by \cite{goli}, whereas this study has managed to show a close approximation even to this wake region by the increased normalized pressure midway when moving from one end to the other. Thus it is quite clear that this study has the capability to predict the results within the vicinity at multiple building heights with reasonable accuracy.\par
\begin{figure}[H]
\centering
\subfloat[Sectional front view]{%
\resizebox*{10cm}{!}{\includegraphics{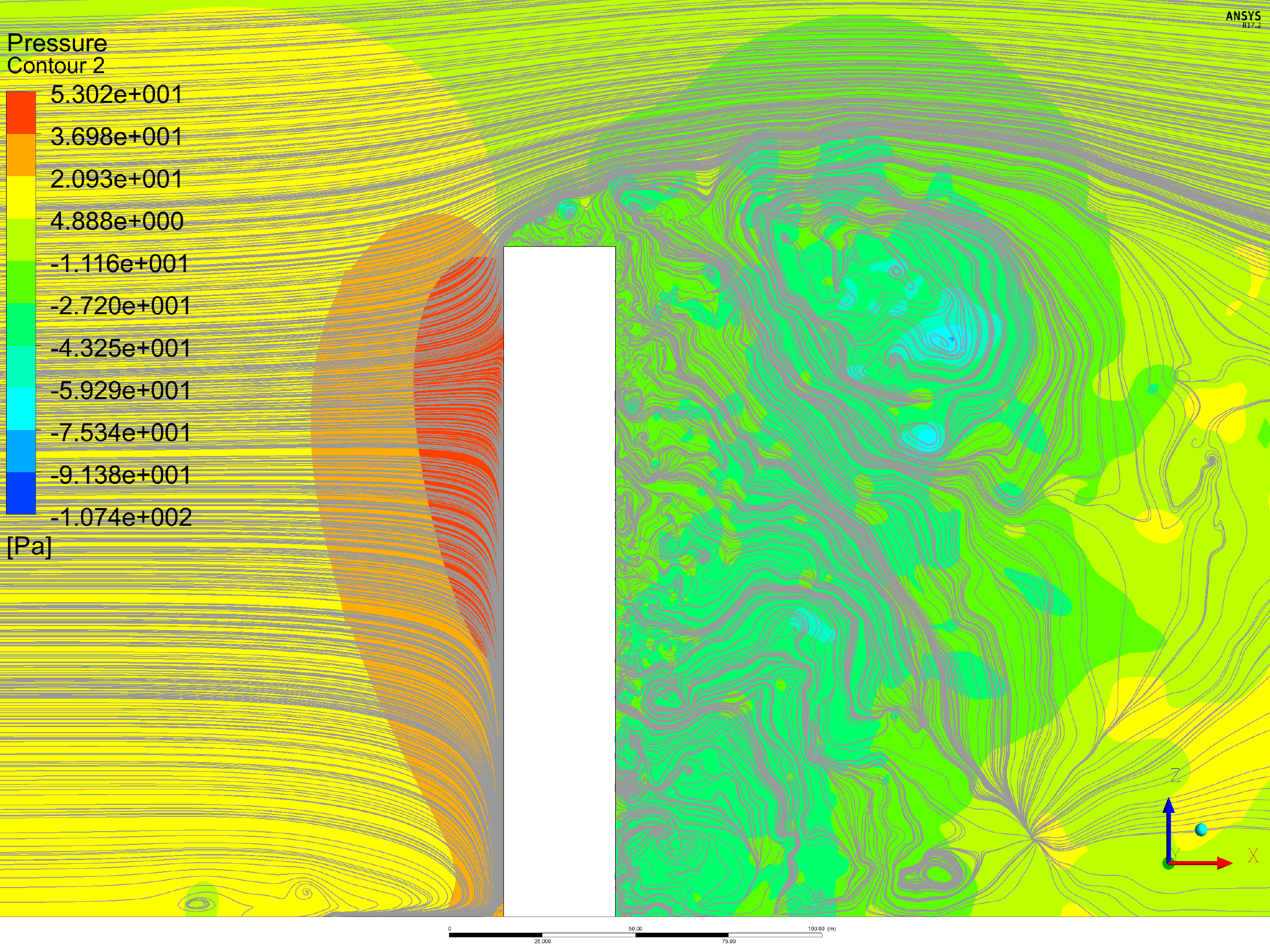}}}\hspace{5pt}
\subfloat[Sectional plan view at building height of 120 m]{%
\resizebox*{10cm}{!}{\includegraphics{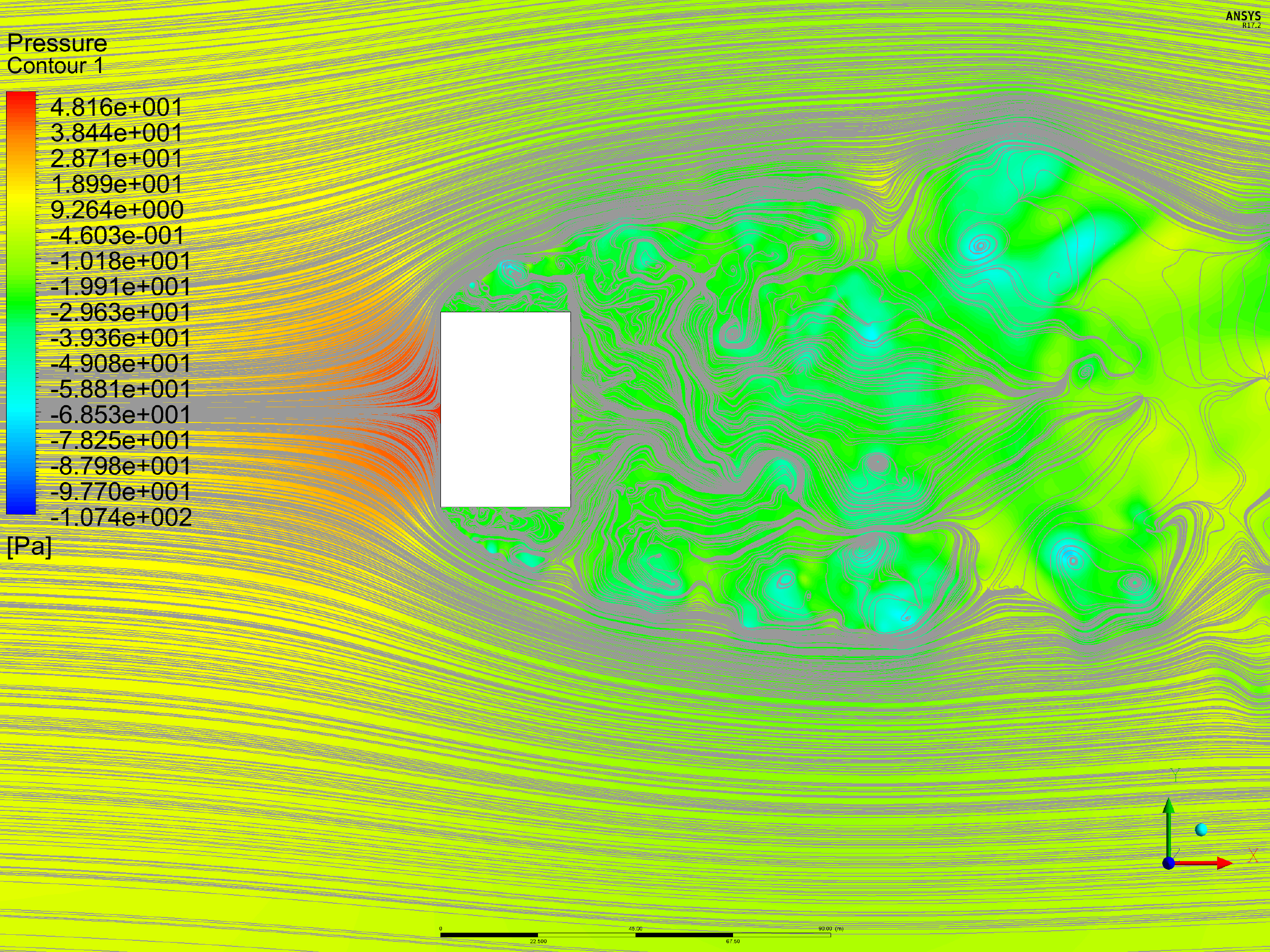}}}
\caption{Velocity streamlines and pressure variation observed for Case 3 } \label{lescoupledview}
\end{figure}

Similar to the velocity and pressure plots in the previous cases, the velocity distribution streamlines and pressure contours were plotted as in Figure  \ref{lescoupledview}. As expected the behaviors obtained closely resembled the observations made in studying Case 2 with respect to the wake formation, wake separation and the suction region. However, the eddy positions seem to be the only variable that could be compared between the two as it is clear that the eddies have shown different displacements which could have been due to the displacements that occur in the fluid mesh as a result of the two-way coupling. Moreover, more eddies are formed at the rear side of the building which is due to the added flexibility of the building.

\subsubsection{Aeroelastic response}

\begin{figure}[H]
\centering
{\includegraphics[scale=0.8]{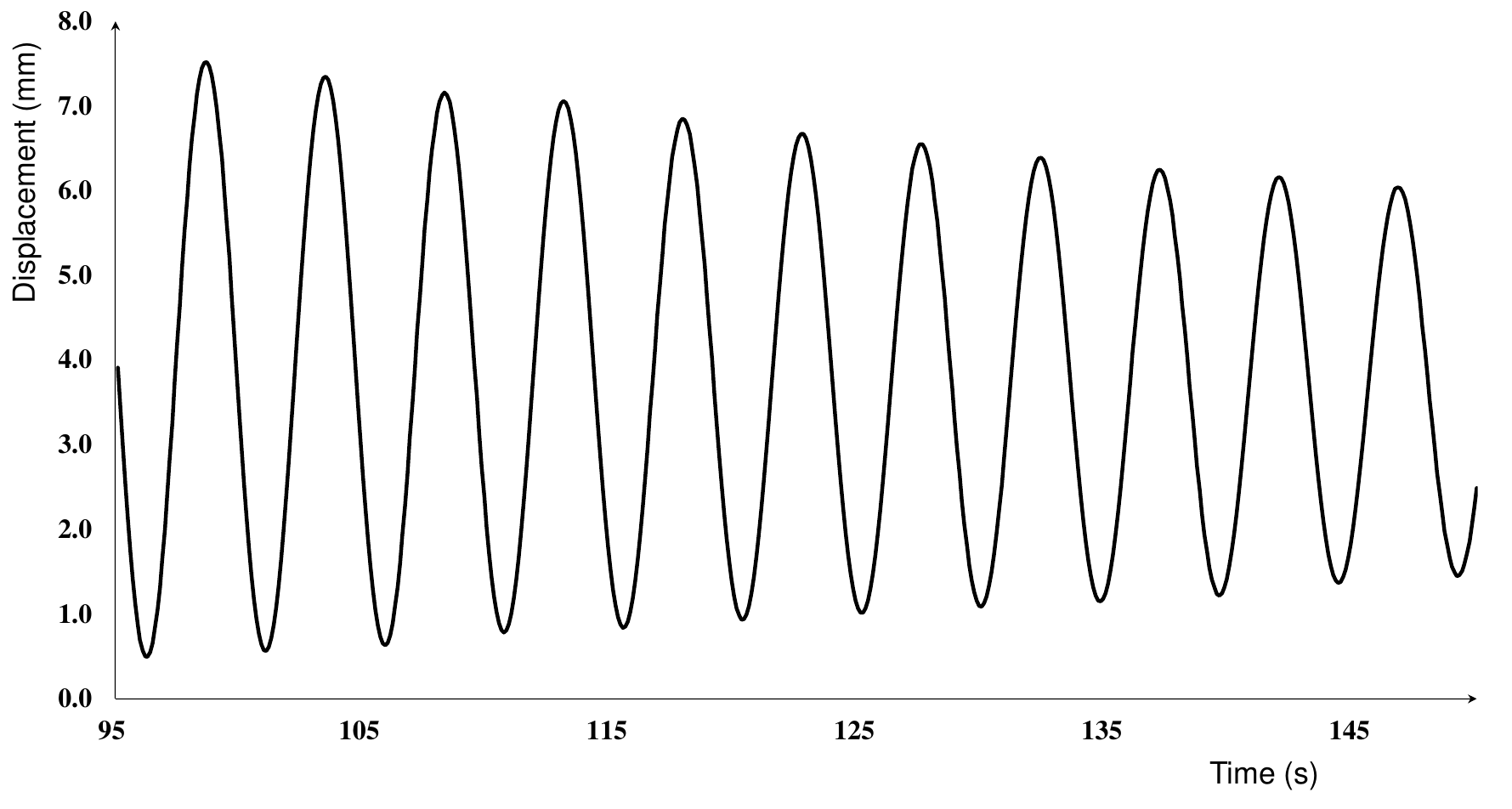}}\hspace{5pt}
\caption{Damped along-wind directional displacement of the building top } \label{alongwind}
\end{figure}

To study the aeroelastic nature the directional wind displacements of the building are studied in along-wind and across-wind directions. Figure \ref{alongwind} shows the variation in displacement of the top of the building (maximum displacement) along the direction of wind propagation which corresponds to the mode shape of the fundamental frequency. According to the variation it is clear that the building vibrations are undergoing sinusoidal displacements with time about a mean value of 4.5 mm with decreasing amplitude over time. As structural damping is not incorporated into the building structure, the main cause for this would be the aeroelastic damping caused in the fluid eddies. The displacement observed is completely positive due to the net pressure force acting on the building as a result of the difference in pressure at frontal and rear faces. 

\begin{figure}[H]
\centering
{\includegraphics[scale=0.8]{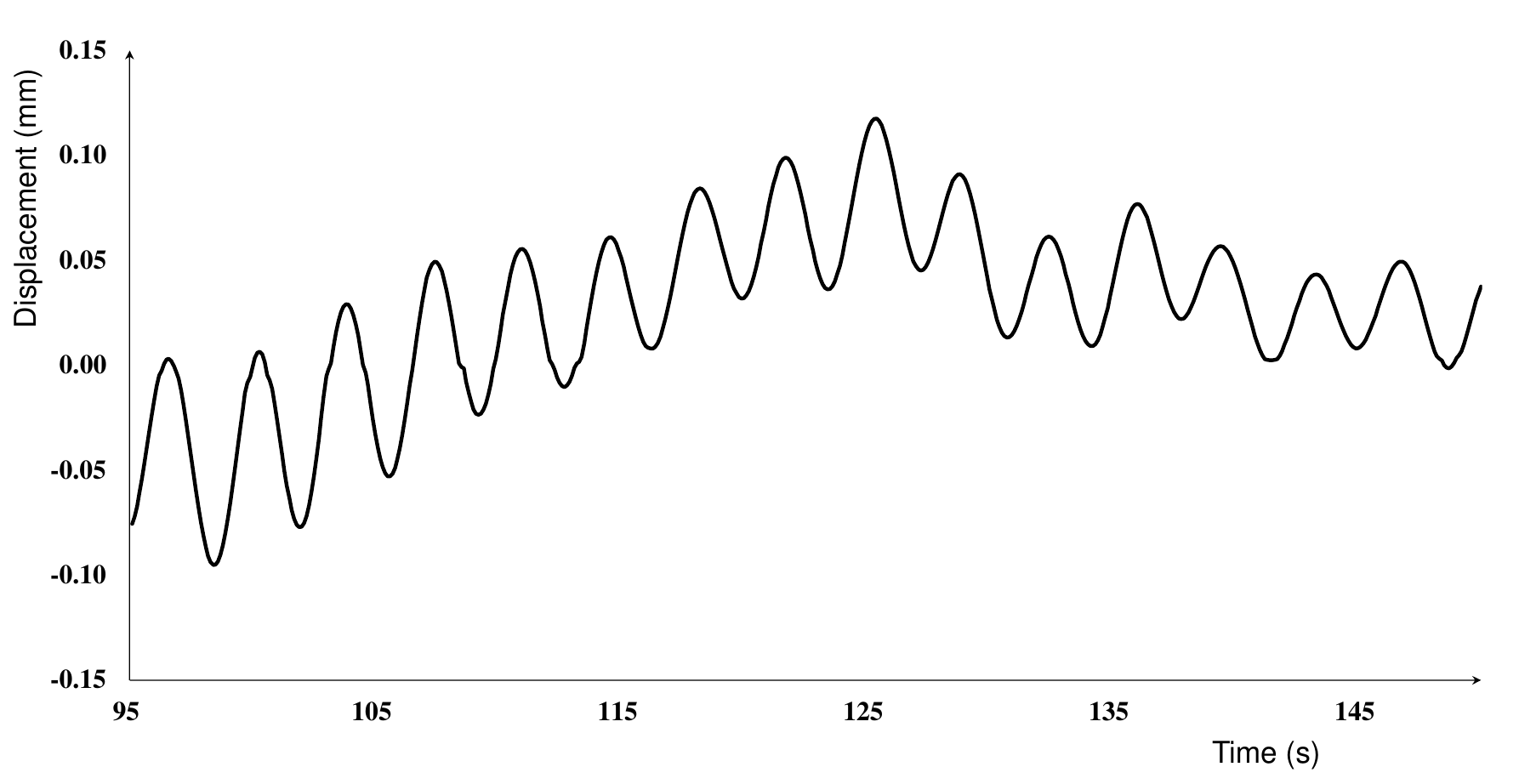}}\hspace{5pt}
\caption{Across-Wind directional displacement of the building top } \label{acrosswind}
\end{figure}

Furthermore, directional displacement was also considered in \emph{y} direction (across-wind response) as shown in Figure \ref{acrosswind}.

\begin{figure}[H]
\centering
{\includegraphics[scale=0.8]{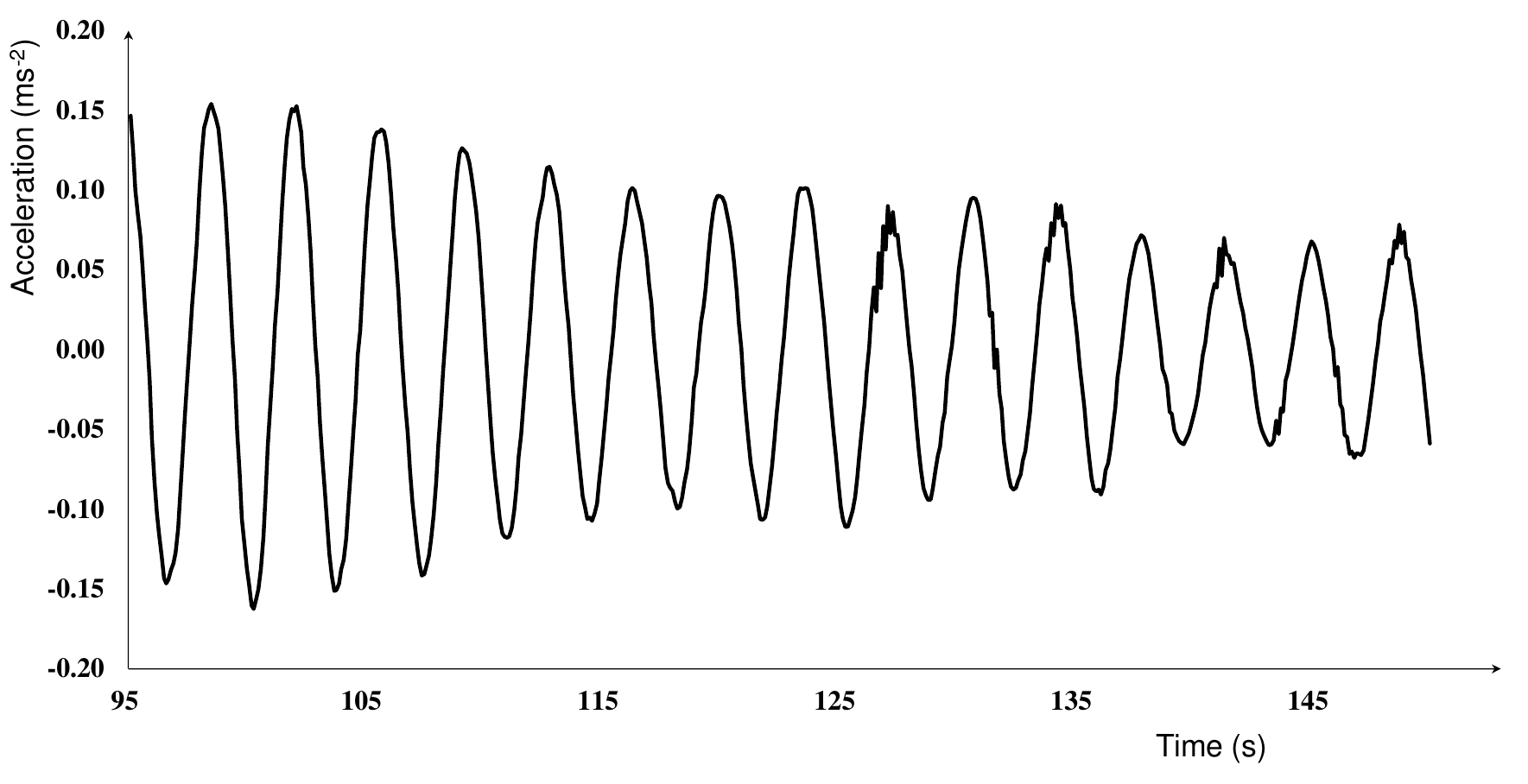}}\hspace{5pt}
\caption{Across-wind directional acceleration of the building top} \label{acrosswindacc}
\end{figure}

The directional deformations do not approach zero according to the obtained results. Furthermore as given in Figure \ref{acrosswindacc} the acceleration of the building in the \emph{y} direction tends to decrease initially but then tend to follow a constant peak to peak variation suggesting an ever present fluctuating force existing in the lateral direction. Thus a fluctuating across-wind response is obtained that lies completely within the positive \emph{y} axis. This is mainly due to the turbulence prediction that is given by the Smagorinsky model. As unbalanced eddies are formed in the two sides a net force is produced towards the lateral direction thus vibrating the building back and forth in addition to the wind induced pressure along \emph{y} direction. The combined effect of these forces in perpendicular directions would result even in moments that would induce torsional vibrations to the building.

\subsection{Comparison}

\begin{figure}[H]
\centering
{\includegraphics[scale=0.8]{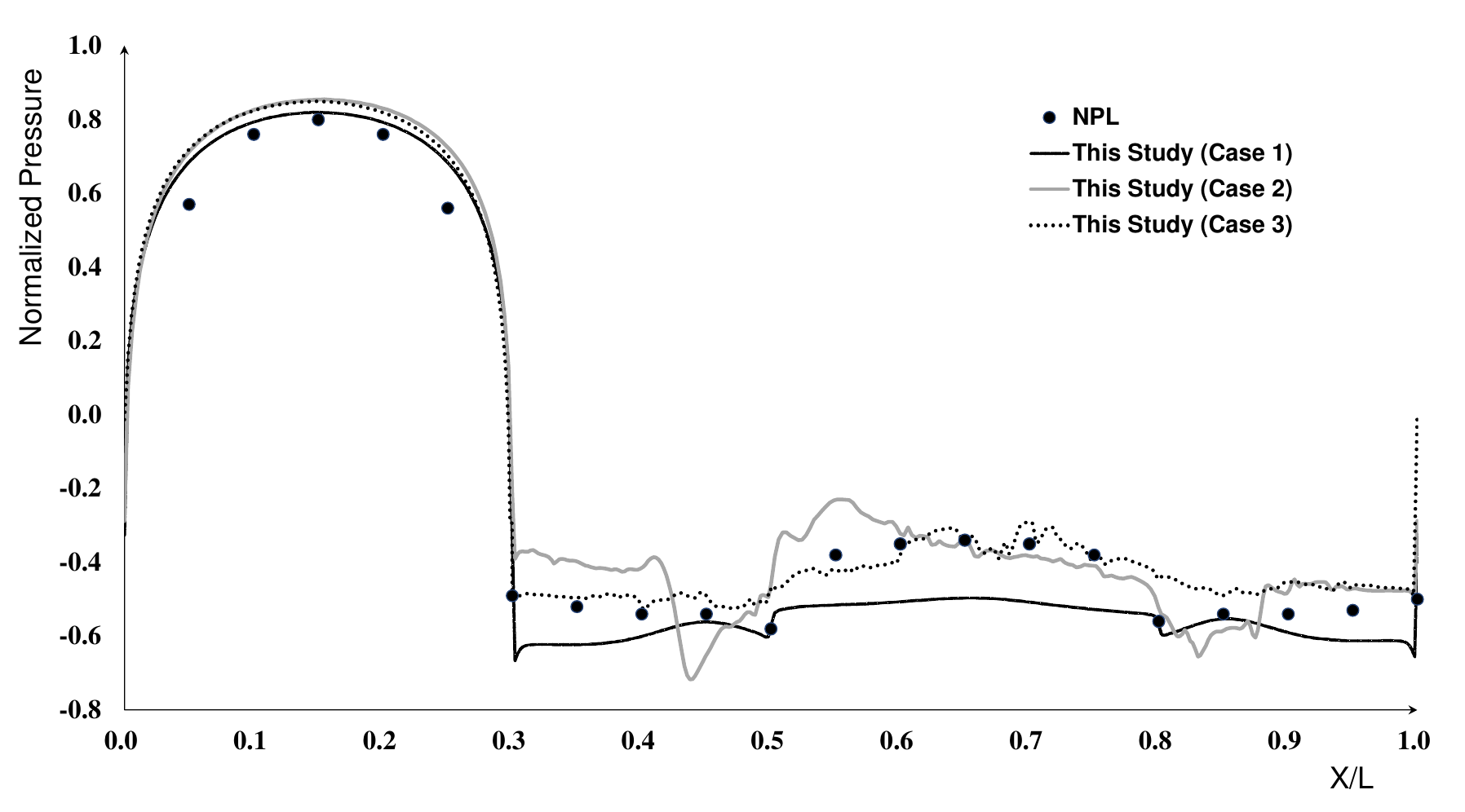}}\hspace{5pt}
\caption{Plot of Normalized pressure along the building perimeter for Case 1, 2 and 3 at building height of 120 m} \label{dis120}
\end{figure}

This section in the paper is utilized to discuss the bottom line of the findings taken from the three cases studied. Since only the NPL study in \cite{mel} has boundary conditions which is closest to this study, the discussion was carried out comparing the normalized pressure at building heights of 120 m and 60 m respectively to the findings of NPL study. However, the comparison at 1/3\textsuperscript{rd} of the height of the building was done using only \cite{goli} since much data is not available for this parameter.The results from all the cases are plotted for comparison purposes for 120 m of building height as in Figure \ref{dis120}.

\begin{figure}[H]
\centering
{\includegraphics[scale=0.8]{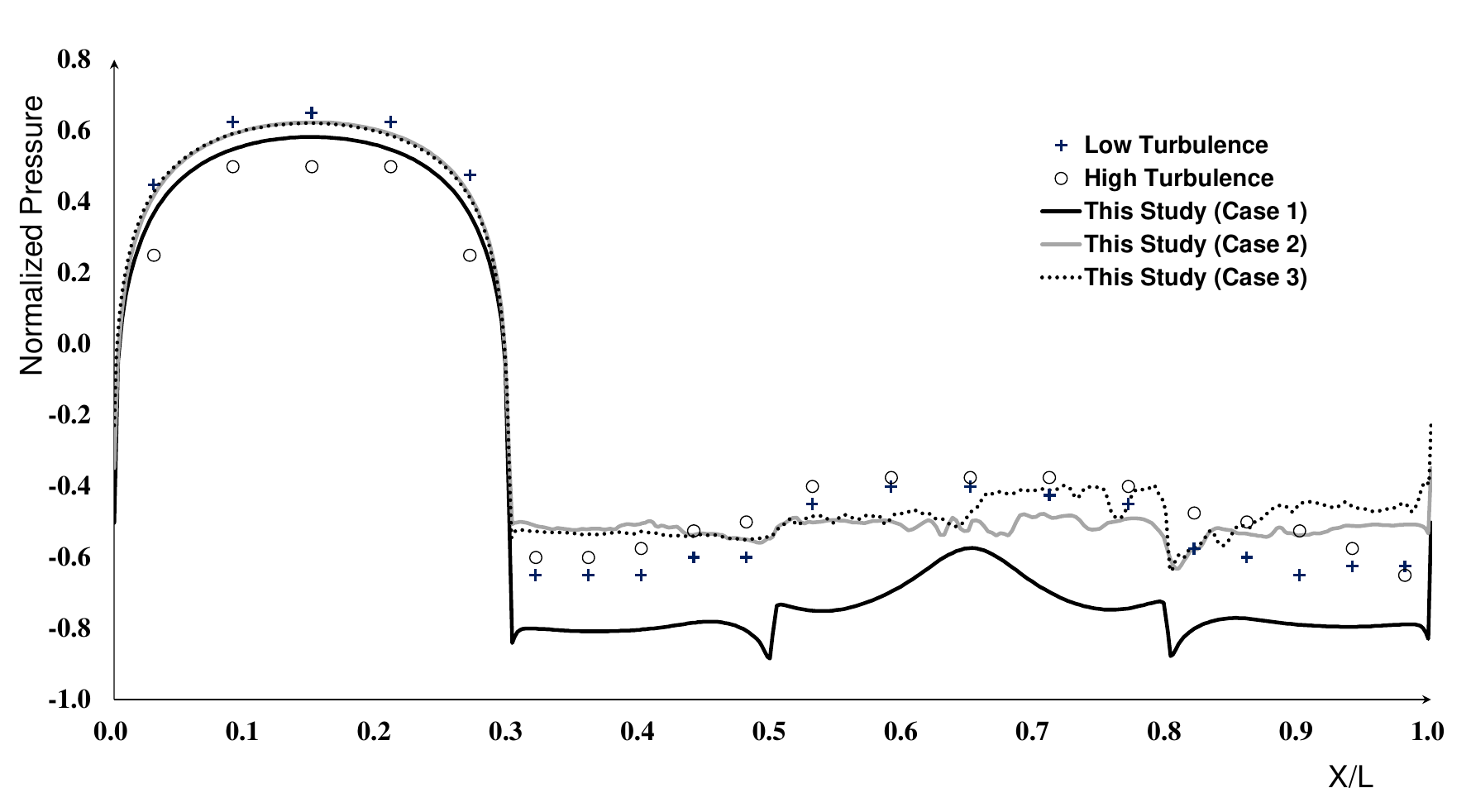}}\hspace{5pt}
\caption{ Plot of Normalized pressure along the building perimeter for Case 1, 2 and 3 at building height of 60 m} \label{dis60}
\end{figure}

The three curves plotted in the same axes pave the way to obtain a comparison with all three cases. In Figure \ref{dis120}, the least difference among the cases is obtained on the front face of the building. This should obviously be the case as such fluid impingement on a wall could only cause a change in pressures which cause vibrations to the building. The behavior shown implies that at the building front face \emph{k}-\textepsilon\ model acting on a rigid building domain is accurate enough to generate an accurate pressure prediction. However, when considering the regions where wake generation is involved it is clearly seen that RANS modeling on a rigid structure would not provide acceptable results. Due to the bending nature that is instilled in the building the aeroelastic model shows very minute deviations relative to the other two involved cases suggesting that once there is a flexible two-way coupling data transfer is efficiently transferred through two models.\par

A similar comparison result was seen in Figure \ref{dis60} as well where the simulation results are compared to experimental values obtained at 1/3\textsuperscript{rd} of building height. Here the \emph{k}-\textepsilon\ model showed a large deviation to the side and rear faces of the building, whereas the rigid model that incorporated an LES turbulence model was not able to show an elevated pressure at the rear face compared to the side faces.However, this was not the case for the coupled analysis, as it has been able to match its pressure variation with experimental results. \par

\begin{figure}[H]
\centering
{\includegraphics[scale=0.8]{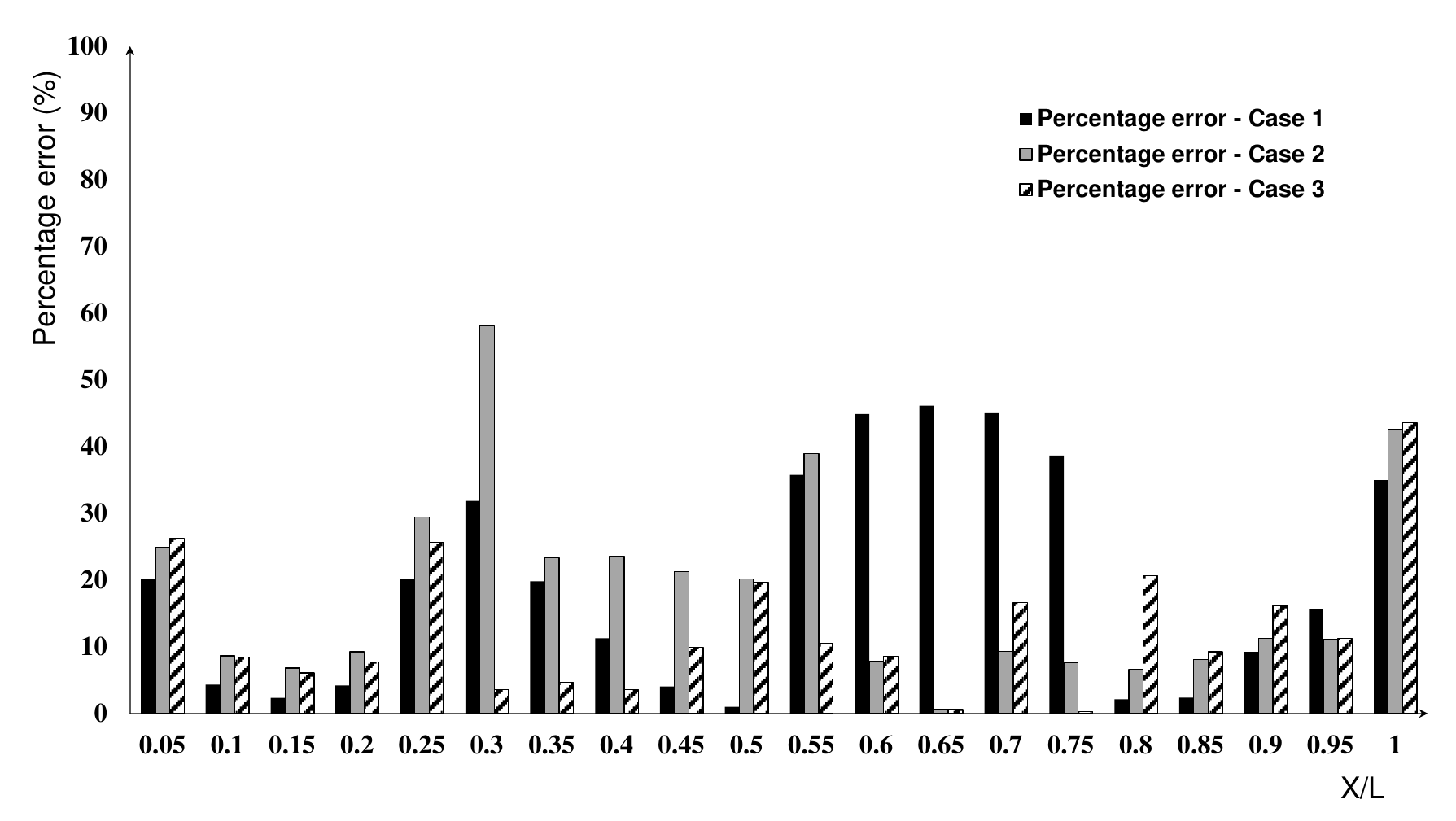}}\hspace{5pt}
\caption{Plot of Normalized pressure along the building perimeter for Case 1, 2 and 3 at building height of 60 m} \label{per}
\end{figure}

To further elaborate the accuracy of the cases discussed a quantitative comparison is done for the percentage deviation of the simulation results from NPL results at building height of 120 m  as shown in Figure \ref{per}.

Figure \ref{per} shows that percentage error in Case 1 has been below 30\% in all the faces except for the rear face where flow separation and wake formation occurs. At the rear face, error percentages have increased even beyond 40\% showing its inability to predict the wake accurately. Case 2 has shown acceptable results, except at the corners where the error percentages have plunged even beyond 50\%. In contrast to both these cases, percentage errors of Case 3 have been less than 20\% for almost all the experimental results and have been significantly less at the corner points due to the building flexibility. Overall, average percentage errors of the cases around the building have mounted to 19.7\%, 18.5\%, 12.7\%  for Cases 1,2 and 3 respectively, whereas, once the front face is excluded the percentages have been reported as 22.2\%, 16.6\%, and 12.5\% respectively suggesting the superiority of the coupled analysis.\par

Furthermore, important observations were made on behalf of the velocity streamlines and wake formation due to the difference in pressure. All three buildings tended to show flow separating in two directions upon striking the building wall. However,  discrepancies were looked upon with respect to the wake formation and the vortice separation in the three cases. The RANS model showed an incomplete wake formation behind the building filled with large eddies. This portrayed the inability of the standard \emph{k}-\textepsilon\ model to capture the wake efficiently even when large Reynolds number flows are employed for which the\emph{k}-\textepsilon\ model with standard wall function is decided as the most optimum method.  In contrast, both Smagorinsky models were able to capture the full effect of wake formation and the final two models differed only through the eddy positions that resulted due to the structural displacement. To discuss the accuracy of the current study further and to stretch its limits beyond the scope of measuring the pressure at a single building height, the pressure at 1/3\textsuperscript{rd} building height was analyzed, and its results did not deviate from the results as in any case in the literature. By the aeroelastic analysis carried out it was verified that the displacements in along-wind direction tend to dominate due to direct wind loads acting on the building, but due to the aeroelastic damping of the wind the building tend to die out its high amplitude vibrations to the mean value in the long term. More importantly, displacements were also observed in the lateral directions where forces would be borne due to the difference in eddy patterns; and convincingly these displacements did not show a consistent damping behavior that would result in a permanent acceleration in the lateral direction.\par

\section{Conclusion}
This work focused on creating an accurate numerical simulation setup for the purpose of studying the wind flow around tall buildings, which has been previously analyzed using various numerical and experimental techniques. Main goal of the study was to devise the setup which would duplicate the same results given by an experimental setup performed at identical conditions; given that available computational resources are used at an optimum level. Comparisons were drawn between different cases for the same CFD grid with respect to the viscous model for fluid flow and nature of the structural domain. An efficient mesh method was introduced with structured blocks having various element sizes and which had tetrahedral, hexahedral mostly and polyhedral elements at locations which had grid size transitions . When the building model was considered to be aeroelastic, the structure was designed to comprise of components with dimensions which would emulate the resonance frequencies of the actual building. Furthermore, the standard system coupling algorithms in FLUENT were used for the aeroelastic analysis as a partitioned coupling method for the first time in a wind engineering application. Important findings obtained from this research could be postulated as follows.
\begin{itemize}
\item The pressure approximation around the building is highly dependent on the nature of the velocity profile and the turbulent intensity profile used. Therefore, always care should be taken to compare the simulation results with experiment results obtained at similar boundary conditions to get a reliable comparison. 

\item The frontal face of the building would provide accurate pressure predictions independent of the turbulence model and the nature of the structural domain. At each of the cases the highest pressure was observed at the frontal face of the building where the fluid flow separated into both upward and downward directions. Therefore, to predict the forces acting on the frontal face minimum computation resources could be used, while the $y$\textsuperscript{+} values stay outside the buffer region. 

\item A Reynolds Averaged method such as the \emph{k}-\textepsilon\ model would not be able to resolve a large percentage of the turbulent kinetic energy. The resulting flow would create two large eddies which are symmetric along the central axis of the building passing perpendicular to the sectional plan view. Thus this symmetricity produces an even pressure distribution in the two shorter sides of the building. Moreover, the normalized pressure distribution in faces other than the frontal face could not show comparable agreement to experimental results for the building heights discussed, especially at the rear face suggesting RANS modeling would not capture the wake region accurately.

\item The use of the Smagorinsky- Lilly model resulted in a more complete wake generation and thus consisted of eddies of much smaller sizes. Further, no symmetricity was observed around any axes for eddy patterns and pressure variations. A much better correlation was observed with experimental results for the wake region suggesting the accuracy of the LES simulations. However,  small inaccuracies are caused due to the displacements that happen in the structural domain which are not taken into account. However , a rigid LES model is sufficient to extract results of a model which would not show substantial structural displacements.
\item The coupled analysis could not show considerable changes in eddy patterns relative to the rigid case as the deformations do not occur at a large scale. However, when the pressure was observed it showed better correlation than the rigid case and results almost overlapped with experimental results. Therefore this method is suitable to fine tune the results obtained for a LES based rigid model study to increase the accuracy. However, this comes at the expense of additional computational time due to the check for convergence that happens for each time step for the data transfer between the two domains.

\item If structural damping is not taken into consideration, a building under wind load would show underdamped sinusoidal displacement in the along-wind direction due to aeroelastic damping. When the wind acts perpendicular to the front face there could not be a across-displacement due to wind load directly. However, due to eddy formation and uneven pressure distribution a net force would be created in the across-wind direction. This was proved to result in a sinusoidal acceleration in the lateral direction.

\end{itemize}
In conclusion, the use of two -way fluid structure interaction is an excellent method to obtain accurate results for wind loads acting on tall structures, when there is sufficient computational power. However, still wind-tunnel tests have an edge over numerical methods in terms of accuracy but both methods are under constant development. This simulation setup could be further improved by doing the study for various wind-incidence angles and also by considering the interference effect caused by neighboring buildings. Also improvements could be made with respect to data transfer and mapping algorithms, grid generation and more reliable subgrid scale turbulence models.

\section*{Acknowledgements}

The authors sincerely acknowledge Department of Mechanical Engineering, University of Moratuwa for providing software facilities for this study.


\begin{thebibliography}{}

\bibitem [\protect \citeauthoryear {%
Aboshosha%
, Alhefnawy%
, Fathelbab%
\BCBL {}\ \BBA {} Shamel%
}{%
Aboshosha%
\ \protect \BOthers {.}}{%
{\protect \APACyear {2010}}%
}]{%
abo}
\APACinsertmetastar {%
abo}%
\begin{APACrefauthors}%
Aboshosha, H.%
, Alhefnawy, L.%
, Fathelbab, F.%
\BCBL {}\ \BBA {} Shamel, A.%
\end{APACrefauthors}%
\unskip\
\newblock
\APACrefYearMonthDay{2010}{}{}.
\newblock
{\BBOQ}\APACrefatitle {Wind Induced Vibrations of Tall Buildings Using CFD}
  {Wind induced vibrations of tall buildings using cfd}.{\BBCQ}
\newblock
\APAChowpublished {The Fifth International Symposium on Computational Wind
  Engineering (CWE2010)}.
\PrintBackRefs{\CurrentBib}

\bibitem [\protect \citeauthoryear {%
Akbay%
, Nobles%
, Zordan%
\BCBL {}\ \BBA {} Shinar%
}{%
Akbay%
\ \protect \BOthers {.}}{%
{\protect \APACyear {2018}}%
}]{%
akbay}
\APACinsertmetastar {%
akbay}%
\begin{APACrefauthors}%
Akbay, M.%
, Nobles, N.%
, Zordan, V.%
\BCBL {}\ \BBA {} Shinar, T.%
\end{APACrefauthors}%
\unskip\
\newblock
\APACrefYearMonthDay{2018}{}{}.
\newblock
{\BBOQ}\APACrefatitle {An Extended Partitioned Method for Conservative
  Solid-Fluid Coupling} {An extended partitioned method for conservative
  solid-fluid coupling}.{\BBCQ}
\newblock
\APACjournalVolNumPages{}{37}{4}{}.
\newblock
\begin{APACrefURL} \url{https://doi.org/10.1145/3197517.3201345}
  \end{APACrefURL}
\newblock
\begin{APACrefDOI} \doi{10.1145/3197517.3201345} \end{APACrefDOI}
\PrintBackRefs{\CurrentBib}

\bibitem [\protect \citeauthoryear {%
Ansys%
}{%
Ansys%
}{%
{\protect \APACyear {2011}}%
}]{%
ansth}
\APACinsertmetastar {%
ansth}%
\begin{APACrefauthors}%
Ansys, I.%
\end{APACrefauthors}%
\unskip\
\newblock
\APACrefYearMonthDay{2011}{}{}.
\newblock
{\BBOQ}\APACrefatitle {ANSYS FLUENT theory guide} {Ansys fluent theory
  guide}.{\BBCQ}
\newblock
\APACjournalVolNumPages{Canonsburg, Pa}{794}{}{}.
\PrintBackRefs{\CurrentBib}

\bibitem [\protect \citeauthoryear {%
Ansys%
}{%
Ansys%
}{%
{\protect \APACyear {2015}}%
}]{%
anssc}
\APACinsertmetastar {%
anssc}%
\begin{APACrefauthors}%
Ansys, I.%
\end{APACrefauthors}%
\unskip\
\newblock
\APACrefYearMonthDay{2015}{}{}.
\newblock
{\BBOQ}\APACrefatitle {ANSYS System Coupling User’s Guide} {Ansys system
  coupling user’s guide}.{\BBCQ}
\newblock
\APACjournalVolNumPages{Canonsburg, Pa}{}{}{}.
\PrintBackRefs{\CurrentBib}

\bibitem [\protect \citeauthoryear {%
Balafas%
}{%
Balafas%
}{%
{\protect \APACyear {2014}}%
}]{%
bala}
\APACinsertmetastar {%
bala}%
\begin{APACrefauthors}%
Balafas, G.%
\end{APACrefauthors}%
\unskip\
\newblock
\APACrefYearMonthDay{2014}{}{}.
\newblock
{\BBOQ}\APACrefatitle {Polyhedral mesh generation for CFD-analysis of complex
  structures} {Polyhedral mesh generation for cfd-analysis of complex
  structures}.{\BBCQ}
\newblock
\APACjournalVolNumPages{Diplomity{\"o}. M{\"u}nchenin teknillinen
  yliopisto}{}{}{}.
\PrintBackRefs{\CurrentBib}

\bibitem [\protect \citeauthoryear {%
Bathe%
\ \BBA {} Zhang%
}{%
Bathe%
\ \BBA {} Zhang%
}{%
{\protect \APACyear {2004}}%
}]{%
bath}
\APACinsertmetastar {%
bath}%
\begin{APACrefauthors}%
Bathe, K\BHBI J.%
\BCBT {}\ \BBA {} Zhang, H.%
\end{APACrefauthors}%
\unskip\
\newblock
\APACrefYearMonthDay{2004}{}{}.
\newblock
{\BBOQ}\APACrefatitle {Finite element developments for general fluid flows with
  structural interactions} {Finite element developments for general fluid flows
  with structural interactions}.{\BBCQ}
\newblock
\APACjournalVolNumPages{International journal for numerical methods in
  engineering}{60}{1}{213--232}.
\PrintBackRefs{\CurrentBib}

\bibitem [\protect \citeauthoryear {%
Blocken%
, Carmeliet%
\BCBL {}\ \BBA {} Stathopoulos%
}{%
Blocken%
\ \protect \BOthers {.}}{%
{\protect \APACyear {2007}}%
}]{%
blocken}
\APACinsertmetastar {%
blocken}%
\begin{APACrefauthors}%
Blocken, B.%
, Carmeliet, J.%
\BCBL {}\ \BBA {} Stathopoulos, T.%
\end{APACrefauthors}%
\unskip\
\newblock
\APACrefYearMonthDay{2007}{}{}.
\newblock
{\BBOQ}\APACrefatitle {CFD evaluation of wind speed conditions in passages
  between parallel buildings—effect of wall-function roughness modifications
  for the atmospheric boundary layer flow} {Cfd evaluation of wind speed
  conditions in passages between parallel buildings—effect of wall-function
  roughness modifications for the atmospheric boundary layer flow}.{\BBCQ}
\newblock
\APACjournalVolNumPages{Journal of Wind Engineering and Industrial
  Aerodynamics}{95}{9-11}{941--962}.
\PrintBackRefs{\CurrentBib}

\bibitem [\protect \citeauthoryear {%
Braun%
\ \BBA {} Awruch%
}{%
Braun%
\ \BBA {} Awruch%
}{%
{\protect \APACyear {2009}}%
}]{%
braun}
\APACinsertmetastar {%
braun}%
\begin{APACrefauthors}%
Braun, A\BPBI L.%
\BCBT {}\ \BBA {} Awruch, A\BPBI M.%
\end{APACrefauthors}%
\unskip\
\newblock
\APACrefYearMonthDay{2009}{}{}.
\newblock
{\BBOQ}\APACrefatitle {Aerodynamic and aeroelastic analyses on the CAARC
  standard tall building model using numerical simulation} {Aerodynamic and
  aeroelastic analyses on the caarc standard tall building model using
  numerical simulation}.{\BBCQ}
\newblock
\APACjournalVolNumPages{Computers \& Structures}{87}{9-10}{564--581}.
\PrintBackRefs{\CurrentBib}

\bibitem [\protect \citeauthoryear {%
Chen%
, Huang%
, Tse%
, Xu%
\BCBL {}\ \BBA {} Li%
}{%
Chen%
, Huang%
\BCBL {}\ \protect \BOthers {.}}{%
{\protect \APACyear {2020}}%
}]{%
chen}
\APACinsertmetastar {%
chen}%
\begin{APACrefauthors}%
Chen, Z.%
, Huang, H.%
, Tse, K.%
, Xu, Y.%
\BCBL {}\ \BBA {} Li, C\BPBI Y.%
\end{APACrefauthors}%
\unskip\
\newblock
\APACrefYearMonthDay{2020}{}{}.
\newblock
{\BBOQ}\APACrefatitle {Characteristics of unsteady aerodynamic forces on an
  aeroelastic prism: A comparative study} {Characteristics of unsteady
  aerodynamic forces on an aeroelastic prism: A comparative study}.{\BBCQ}
\newblock
\APACjournalVolNumPages{Journal of Wind Engineering and Industrial
  Aerodynamics}{205}{}{104325}.
\newblock
\begin{APACrefDOI} \doi{https://doi.org/10.1016/j.jweia.2020.104325}
  \end{APACrefDOI}
\PrintBackRefs{\CurrentBib}

\bibitem [\protect \citeauthoryear {%
Chen%
, Xu%
, Huang%
\BCBL {}\ \BBA {} Tse%
}{%
Chen%
, Xu%
\BCBL {}\ \protect \BOthers {.}}{%
{\protect \APACyear {2020}}%
}]{%
zeng20}
\APACinsertmetastar {%
zeng20}%
\begin{APACrefauthors}%
Chen, Z.%
, Xu, Y.%
, Huang, H.%
\BCBL {}\ \BBA {} Tse, K\BPBI T.%
\end{APACrefauthors}%
\unskip\
\newblock
\APACrefYearMonthDay{2020}{}{}.
\newblock
{\BBOQ}\APACrefatitle {Wind tunnel measurement systems for unsteady aerodynamic
  forces on bluff bodies: review and new perspective} {Wind tunnel measurement
  systems for unsteady aerodynamic forces on bluff bodies: review and new
  perspective}.{\BBCQ}
\newblock
\APACjournalVolNumPages{Sensors}{20}{16}{4633}.
\PrintBackRefs{\CurrentBib}

\bibitem [\protect \citeauthoryear {%
Chimakurthi%
, Reuss%
, Tooley%
\BCBL {}\ \BBA {} Scampoli%
}{%
Chimakurthi%
\ \protect \BOthers {.}}{%
{\protect \APACyear {1995}}%
}]{%
chima}
\APACinsertmetastar {%
chima}%
\begin{APACrefauthors}%
Chimakurthi, S.%
, Reuss, S.%
, Tooley, M.%
\BCBL {}\ \BBA {} Scampoli, S.%
\end{APACrefauthors}%
\unskip\
\newblock
\APACrefYearMonthDay{1995}{}{}.
\newblock
{\BBOQ}\APACrefatitle {ANSYS Workbench System Coupling: a state-of-the-art
  computational framework for analyzing multiphysics problems} {Ansys workbench
  system coupling: a state-of-the-art computational framework for analyzing
  multiphysics problems}.{\BBCQ}
\newblock

\PrintBackRefs{\CurrentBib}

\bibitem [\protect \citeauthoryear {%
Dagnew%
, Bitsuamalk%
\BCBL {}\ \BBA {} Merrick%
}{%
Dagnew%
\ \protect \BOthers {.}}{%
{\protect \APACyear {2009}}%
}]{%
dagnew}
\APACinsertmetastar {%
dagnew}%
\begin{APACrefauthors}%
Dagnew, A\BPBI K.%
, Bitsuamalk, G\BPBI T.%
\BCBL {}\ \BBA {} Merrick, R.%
\end{APACrefauthors}%
\unskip\
\newblock
\APACrefYearMonthDay{2009}{}{}.
\newblock
{\BBOQ}\APACrefatitle {Computational evaluation of wind pressures on tall
  buildings} {Computational evaluation of wind pressures on tall
  buildings}.{\BBCQ}
\newblock
\BIn{} \APACrefbtitle {11th American conference on Wind Engineering. San Juan,
  Puerto Rico.} {11th american conference on wind engineering. san juan, puerto
  rico.}
\PrintBackRefs{\CurrentBib}

\bibitem [\protect \citeauthoryear {%
Galpin%
}{%
Galpin%
}{%
{\protect \APACyear {1995}}%
}]{%
galpin}
\APACinsertmetastar {%
galpin}%
\begin{APACrefauthors}%
Galpin, H., Broberg.%
\end{APACrefauthors}%
\unskip\
\newblock
\APACrefYearMonthDay{1995}{}{}.
\newblock
{\BBOQ}\APACrefatitle {Three Dimensional Navier Stokes Predictions of Steady
  State Rotor} {Three dimensional navier stokes predictions of steady state
  rotor}.{\BBCQ}
\newblock

\PrintBackRefs{\CurrentBib}

\bibitem [\protect \citeauthoryear {%
Goliger%
\ \BBA {} Milford%
}{%
Goliger%
\ \BBA {} Milford%
}{%
{\protect \APACyear {1988}}%
}]{%
goli}
\APACinsertmetastar {%
goli}%
\begin{APACrefauthors}%
Goliger, A.%
\BCBT {}\ \BBA {} Milford, R.%
\end{APACrefauthors}%
\unskip\
\newblock
\APACrefYearMonthDay{1988}{}{}.
\newblock
{\BBOQ}\APACrefatitle {Sensitivity of the CAARC standard building model to
  geometric scale and turbulence} {Sensitivity of the caarc standard building
  model to geometric scale and turbulence}.{\BBCQ}
\newblock
\APACjournalVolNumPages{Journal of Wind Engineering and Industrial
  Aerodynamics}{31}{1}{105--123}.
\PrintBackRefs{\CurrentBib}

\bibitem [\protect \citeauthoryear {%
Hanson%
, Smith%
, Summers%
\BCBL {}\ \BBA {} Wilson%
}{%
Hanson%
\ \protect \BOthers {.}}{%
{\protect \APACyear {1982}}%
}]{%
hans82}
\APACinsertmetastar {%
hans82}%
\begin{APACrefauthors}%
Hanson, T.%
, Smith, F.%
, Summers, D.%
\BCBL {}\ \BBA {} Wilson, C.%
\end{APACrefauthors}%
\unskip\
\newblock
\APACrefYearMonthDay{1982}{}{}.
\newblock
{\BBOQ}\APACrefatitle {Computer simulation of wind flow around buildings}
  {Computer simulation of wind flow around buildings}.{\BBCQ}
\newblock
\APACjournalVolNumPages{Computer-Aided Design}{14}{1}{27--31}.
\PrintBackRefs{\CurrentBib}

\bibitem [\protect \citeauthoryear {%
Hanson%
, Summers%
\BCBL {}\ \BBA {} Wilson%
}{%
Hanson%
\ \protect \BOthers {.}}{%
{\protect \APACyear {1986}}%
}]{%
hans86}
\APACinsertmetastar {%
hans86}%
\begin{APACrefauthors}%
Hanson, T.%
, Summers, D.%
\BCBL {}\ \BBA {} Wilson, C.%
\end{APACrefauthors}%
\unskip\
\newblock
\APACrefYearMonthDay{1986}{}{}.
\newblock
{\BBOQ}\APACrefatitle {A three-dimensional simulation of wind flow around
  buildings} {A three-dimensional simulation of wind flow around
  buildings}.{\BBCQ}
\newblock
\APACjournalVolNumPages{International journal for numerical methods in
  fluids}{6}{3}{113--127}.
\PrintBackRefs{\CurrentBib}

\bibitem [\protect \citeauthoryear {%
Hirt%
}{%
Hirt%
}{%
{\protect \APACyear {1975}}%
}]{%
hirt75}
\APACinsertmetastar {%
hirt75}%
\begin{APACrefauthors}%
Hirt, C.%
\end{APACrefauthors}%
\unskip\
\newblock
\APACrefYearMonthDay{1975}{}{}.
\newblock
{\BBOQ}\APACrefatitle {A numerical solution algorithm for transient fluid flow}
  {A numerical solution algorithm for transient fluid flow}.{\BBCQ}
\newblock
\APACjournalVolNumPages{Los Alamos Scientific Laboratory Report}{}{}{}.
\PrintBackRefs{\CurrentBib}

\bibitem [\protect \citeauthoryear {%
Hirt%
, Ramshaw%
\BCBL {}\ \BBA {} Stein%
}{%
Hirt%
\ \protect \BOthers {.}}{%
{\protect \APACyear {1978}}%
}]{%
hirt}
\APACinsertmetastar {%
hirt}%
\begin{APACrefauthors}%
Hirt, C.%
, Ramshaw, J.%
\BCBL {}\ \BBA {} Stein, L.%
\end{APACrefauthors}%
\unskip\
\newblock
\APACrefYearMonthDay{1978}{}{}.
\newblock
{\BBOQ}\APACrefatitle {Numerical simulation of three-dimensional flow past
  bluff bodies} {Numerical simulation of three-dimensional flow past bluff
  bodies}.{\BBCQ}
\newblock
\APACjournalVolNumPages{Computer Methods in Applied Mechanics and
  Engineering}{14}{1}{93--124}.
\PrintBackRefs{\CurrentBib}

\bibitem [\protect \citeauthoryear {%
Huang%
, Li%
\BCBL {}\ \BBA {} Xu%
}{%
Huang%
\ \protect \BOthers {.}}{%
{\protect \APACyear {2007}}%
}]{%
huang}
\APACinsertmetastar {%
huang}%
\begin{APACrefauthors}%
Huang, S.%
, Li, Q.%
\BCBL {}\ \BBA {} Xu, S.%
\end{APACrefauthors}%
\unskip\
\newblock
\APACrefYearMonthDay{2007}{}{}.
\newblock
{\BBOQ}\APACrefatitle {Numerical evaluation of wind effects on a tall steel
  building by CFD} {Numerical evaluation of wind effects on a tall steel
  building by cfd}.{\BBCQ}
\newblock
\APACjournalVolNumPages{Journal of Constructional Steel
  Research}{63}{5}{612--627}.
\PrintBackRefs{\CurrentBib}

\bibitem [\protect \citeauthoryear {%
{\.I}rsel%
}{%
{\.I}rsel%
}{%
{\protect \APACyear {2019}}%
}]{%
irsel}
\APACinsertmetastar {%
irsel}%
\begin{APACrefauthors}%
{\.I}rsel, G.%
\end{APACrefauthors}%
\unskip\
\newblock
\APACrefYearMonthDay{2019}{}{}.
\newblock
{\BBOQ}\APACrefatitle {The effect of using shell and solid models in structural
  stress analysis} {The effect of using shell and solid models in structural
  stress analysis}.{\BBCQ}
\newblock
\APACjournalVolNumPages{Vibroengineering PROCEDIA}{27}{}{115--120}.
\PrintBackRefs{\CurrentBib}

\bibitem [\protect \citeauthoryear {%
Jansen%
, Shakib%
\BCBL {}\ \BBA {} Hughes%
}{%
Jansen%
\ \protect \BOthers {.}}{%
{\protect \APACyear {1992}}%
}]{%
jansen}
\APACinsertmetastar {%
jansen}%
\begin{APACrefauthors}%
Jansen, K.%
, Shakib, F.%
\BCBL {}\ \BBA {} Hughes, T\BPBI J.%
\end{APACrefauthors}%
\unskip\
\newblock
\APACrefYearMonthDay{1992}{}{}.
\newblock
{\BBOQ}\APACrefatitle {Fast projection algorithm for unstructured meshes} {Fast
  projection algorithm for unstructured meshes}.{\BBCQ}
\newblock

\PrintBackRefs{\CurrentBib}

\bibitem [\protect \citeauthoryear {%
Koliyabandara%
, Jayasundara%
\BCBL {}\ \BBA {} Wijesundara%
}{%
Koliyabandara%
\ \protect \BOthers {.}}{%
{\protect \APACyear {2017}}%
}]{%
koli17}
\APACinsertmetastar {%
koli17}%
\begin{APACrefauthors}%
Koliyabandara, S.%
, Jayasundara, H.%
\BCBL {}\ \BBA {} Wijesundara, K.%
\end{APACrefauthors}%
\unskip\
\newblock
\APACrefYearMonthDay{2017}{}{}.
\newblock
{\BBOQ}\APACrefatitle {ANALYSIS OF WIND LOADS ON AN IRREGULAR SHAPED TALL
  BUILDING USING NUMERICAL SIMULATION} {Analysis of wind loads on an irregular
  shaped tall building using numerical simulation}.{\BBCQ}
\newblock
\APACjournalVolNumPages{ICSECM2017-456}{}{}{}.
\PrintBackRefs{\CurrentBib}

\bibitem [\protect \citeauthoryear {%
Koliyabandara%
, Jayasundara%
\BCBL {}\ \BBA {} Wijesundara%
}{%
Koliyabandara%
\ \protect \BOthers {.}}{%
{\protect \APACyear {2018}}%
}]{%
koli18}
\APACinsertmetastar {%
koli18}%
\begin{APACrefauthors}%
Koliyabandara, S.%
, Jayasundara, H.%
\BCBL {}\ \BBA {} Wijesundara, K.%
\end{APACrefauthors}%
\unskip\
\newblock
\APACrefYearMonthDay{2018}{}{}.
\newblock
{\BBOQ}\APACrefatitle {Evaluation of Different Turbulence Models in Determining
  Wind Loads on Tall Buildings} {Evaluation of different turbulence models in
  determining wind loads on tall buildings}.{\BBCQ}
\newblock

\PrintBackRefs{\CurrentBib}

\bibitem [\protect \citeauthoryear {%
~Launder%
}{%
~Launder%
}{%
{\protect \APACyear {1993}}%
}]{%
lau93}
\APACinsertmetastar {%
lau93}%
\begin{APACrefauthors}%
Launder%
\end{APACrefauthors}%
\unskip\
\newblock
\APACrefYearMonthDay{1993}{}{}.
\newblock
{\BBOQ}\APACrefatitle {Modeling flow-induced oscillations in turbulent flow
  around a square cylinder} {Modeling flow-induced oscillations in turbulent
  flow around a square cylinder}.{\BBCQ}
\newblock
\BIn{} \APACrefbtitle {ASME Fluid Engineering conference 1993.} {Asme fluid
  engineering conference 1993.}
\PrintBackRefs{\CurrentBib}

\bibitem [\protect \citeauthoryear {%
Launder%
\ \BBA {} Sharma%
}{%
Launder%
\ \BBA {} Sharma%
}{%
{\protect \APACyear {1974}}%
}]{%
launder}
\APACinsertmetastar {%
launder}%
\begin{APACrefauthors}%
Launder, B\BPBI E.  {}\ \BBA {} Sharma, B\BPBI I.%
\end{APACrefauthors}%
\unskip\
\newblock
\APACrefYearMonthDay{1974}{}{}.
\newblock
{\BBOQ}\APACrefatitle {Application of the energy-dissipation model of
  turbulence to the calculation of flow near a spinning disc} {Application of
  the energy-dissipation model of turbulence to the calculation of flow near a
  spinning disc}.{\BBCQ}
\newblock
\APACjournalVolNumPages{Letters in heat and mass transfer}{1}{2}{131--137}.
\PrintBackRefs{\CurrentBib}

\bibitem [\protect \citeauthoryear {%
Melbourne%
}{%
Melbourne%
}{%
{\protect \APACyear {1980}}%
}]{%
mel}
\APACinsertmetastar {%
mel}%
\begin{APACrefauthors}%
Melbourne, W.%
\end{APACrefauthors}%
\unskip\
\newblock
\APACrefYearMonthDay{1980}{}{}.
\newblock
{\BBOQ}\APACrefatitle {Comparison of measurements on the CAARC standard tall
  building model in simulated model wind flows} {Comparison of measurements on
  the caarc standard tall building model in simulated model wind flows}.{\BBCQ}
\newblock
\APACjournalVolNumPages{Journal of Wind Engineering and Industrial
  Aerodynamics}{6}{1-2}{73--88}.
\PrintBackRefs{\CurrentBib}

\bibitem [\protect \citeauthoryear {%
Mochida%
\ \BBA {} Lun%
}{%
Mochida%
\ \BBA {} Lun%
}{%
{\protect \APACyear {2008}}%
}]{%
mochida}
\APACinsertmetastar {%
mochida}%
\begin{APACrefauthors}%
Mochida, A.%
\BCBT {}\ \BBA {} Lun, I\BPBI Y.%
\end{APACrefauthors}%
\unskip\
\newblock
\APACrefYearMonthDay{2008}{}{}.
\newblock
{\BBOQ}\APACrefatitle {Prediction of wind environment and thermal comfort at
  pedestrian level in urban area} {Prediction of wind environment and thermal
  comfort at pedestrian level in urban area}.{\BBCQ}
\newblock
\APACjournalVolNumPages{Journal of wind engineering and industrial
  aerodynamics}{96}{10-11}{1498--1527}.
\PrintBackRefs{\CurrentBib}

\bibitem [\protect \citeauthoryear {%
Mou%
, He%
, Zhao%
\BCBL {}\ \BBA {} Chau%
}{%
Mou%
\ \protect \BOthers {.}}{%
{\protect \APACyear {2017}}%
}]{%
mou}
\APACinsertmetastar {%
mou}%
\begin{APACrefauthors}%
Mou, B.%
, He, B\BHBI J.%
, Zhao, D\BHBI X.%
\BCBL {}\ \BBA {} Chau, K\BHBI w.%
\end{APACrefauthors}%
\unskip\
\newblock
\APACrefYearMonthDay{2017}{}{}.
\newblock
{\BBOQ}\APACrefatitle {Numerical simulation of the effects of building
  dimensional variation on wind pressure distribution} {Numerical simulation of
  the effects of building dimensional variation on wind pressure
  distribution}.{\BBCQ}
\newblock
\APACjournalVolNumPages{Engineering Applications of Computational Fluid
  Mechanics}{11}{1}{293--309}.
\PrintBackRefs{\CurrentBib}

\bibitem [\protect \citeauthoryear {%
Murakami%
}{%
Murakami%
}{%
{\protect \APACyear {1993}}%
}]{%
mura93}
\APACinsertmetastar {%
mura93}%
\begin{APACrefauthors}%
Murakami, S.%
\end{APACrefauthors}%
\unskip\
\newblock
\APACrefYearMonthDay{1993}{}{}.
\newblock
{\BBOQ}\APACrefatitle {Comparison of various turbulence models applied to a
  bluff body} {Comparison of various turbulence models applied to a bluff
  body}.{\BBCQ}
\newblock
\BIn{} \APACrefbtitle {Computational Wind Engineering 1} {Computational wind
  engineering 1}\ (\BPGS\ 21--36).
\newblock
\APACaddressPublisher{}{Elsevier}.
\PrintBackRefs{\CurrentBib}

\bibitem [\protect \citeauthoryear {%
Murakami%
}{%
Murakami%
}{%
{\protect \APACyear {1997}}%
}]{%
mura97}
\APACinsertmetastar {%
mura97}%
\begin{APACrefauthors}%
Murakami, S.%
\end{APACrefauthors}%
\unskip\
\newblock
\APACrefYearMonthDay{1997}{}{}.
\newblock
{\BBOQ}\APACrefatitle {Current status and future trends in computational wind
  engineering} {Current status and future trends in computational wind
  engineering}.{\BBCQ}
\newblock
\APACjournalVolNumPages{Journal of wind engineering and industrial
  aerodynamics}{67}{}{3--34}.
\PrintBackRefs{\CurrentBib}

\bibitem [\protect \citeauthoryear {%
Murakami%
}{%
Murakami%
}{%
{\protect \APACyear {1998}}%
}]{%
mura98}
\APACinsertmetastar {%
mura98}%
\begin{APACrefauthors}%
Murakami, S.%
\end{APACrefauthors}%
\unskip\
\newblock
\APACrefYearMonthDay{1998}{}{}.
\newblock
{\BBOQ}\APACrefatitle {Overview of turbulence models applied in CWE--1997}
  {Overview of turbulence models applied in cwe--1997}.{\BBCQ}
\newblock
\APACjournalVolNumPages{Journal of Wind Engineering and Industrial
  Aerodynamics}{74}{}{1--24}.
\PrintBackRefs{\CurrentBib}

\bibitem [\protect \citeauthoryear {%
Murakami%
\ \BBA {} Mochida%
}{%
Murakami%
\ \BBA {} Mochida%
}{%
{\protect \APACyear {1989}}%
}]{%
mura89}
\APACinsertmetastar {%
mura89}%
\begin{APACrefauthors}%
Murakami, S.%
\BCBT {}\ \BBA {} Mochida, A.%
\end{APACrefauthors}%
\unskip\
\newblock
\APACrefYearMonthDay{1989}{}{}.
\newblock
{\BBOQ}\APACrefatitle {Three-dimensional numerical simulation of turbulent flow
  around buildings using the k- $\varepsilon$ turbulence model}
  {Three-dimensional numerical simulation of turbulent flow around buildings
  using the k- $\varepsilon$ turbulence model}.{\BBCQ}
\newblock
\APACjournalVolNumPages{Building and Environment}{24}{1}{51--64}.
\PrintBackRefs{\CurrentBib}

\bibitem [\protect \citeauthoryear {%
Murakami%
, Mochida%
\BCBL {}\ \BBA {} Hayashi%
}{%
Murakami%
\ \protect \BOthers {.}}{%
{\protect \APACyear {1990}}%
}]{%
mura90}
\APACinsertmetastar {%
mura90}%
\begin{APACrefauthors}%
Murakami, S.%
, Mochida, A.%
\BCBL {}\ \BBA {} Hayashi, Y.%
\end{APACrefauthors}%
\unskip\
\newblock
\APACrefYearMonthDay{1990}{}{}.
\newblock
{\BBOQ}\APACrefatitle {Examining the \textkappa\ -\textepsilon\ model by means
  of a wind tunnel test and large-eddy simulation of the turbulence structure
  around a cube} {Examining the \textkappa\ -\textepsilon\ model by means of a
  wind tunnel test and large-eddy simulation of the turbulence structure around
  a cube}.{\BBCQ}
\newblock
\APACjournalVolNumPages{Journal of wind engineering and Industrial
  Aerodynamics}{35}{}{87--100}.
\PrintBackRefs{\CurrentBib}

\bibitem [\protect \citeauthoryear {%
Obasaju%
}{%
Obasaju%
}{%
{\protect \APACyear {1992}}%
}]{%
oba}
\APACinsertmetastar {%
oba}%
\begin{APACrefauthors}%
Obasaju, E.%
\end{APACrefauthors}%
\unskip\
\newblock
\APACrefYearMonthDay{1992}{}{}.
\newblock
{\BBOQ}\APACrefatitle {Measurement of forces and base overturning moments on
  the CAARC tall building model in a simulated atmospheric boundary layer}
  {Measurement of forces and base overturning moments on the caarc tall
  building model in a simulated atmospheric boundary layer}.{\BBCQ}
\newblock
\APACjournalVolNumPages{Journal of Wind Engineering and Industrial
  Aerodynamics}{40}{2}{103--126}.
\PrintBackRefs{\CurrentBib}

\bibitem [\protect \citeauthoryear {%
Ramponi%
\ \BBA {} Blocken%
}{%
Ramponi%
\ \BBA {} Blocken%
}{%
{\protect \APACyear {2012}}%
}]{%
ramponi}
\APACinsertmetastar {%
ramponi}%
\begin{APACrefauthors}%
Ramponi, R.%
\BCBT {}\ \BBA {} Blocken, B.%
\end{APACrefauthors}%
\unskip\
\newblock
\APACrefYearMonthDay{2012}{}{}.
\newblock
{\BBOQ}\APACrefatitle {CFD simulation of cross-ventilation for a generic
  isolated building: impact of computational parameters} {Cfd simulation of
  cross-ventilation for a generic isolated building: impact of computational
  parameters}.{\BBCQ}
\newblock
\APACjournalVolNumPages{Building and environment}{53}{}{34--48}.
\PrintBackRefs{\CurrentBib}

\bibitem [\protect \citeauthoryear {%
Rugonyi%
\ \BBA {} Bathe%
}{%
Rugonyi%
\ \BBA {} Bathe%
}{%
{\protect \APACyear {2001}}%
}]{%
rugo}
\APACinsertmetastar {%
rugo}%
\begin{APACrefauthors}%
Rugonyi, S.%
\BCBT {}\ \BBA {} Bathe, K\BHBI J.%
\end{APACrefauthors}%
\unskip\
\newblock
\APACrefYearMonthDay{2001}{}{}.
\newblock
{\BBOQ}\APACrefatitle {On finite element analysis of fluid flows fully coupled
  with structural interactions} {On finite element analysis of fluid flows
  fully coupled with structural interactions}.{\BBCQ}
\newblock
\APACjournalVolNumPages{CMES- Computer Modeling in Engineering and
  Sciences}{2}{2}{195--212}.
\PrintBackRefs{\CurrentBib}

\bibitem [\protect \citeauthoryear {%
Stathopoulos%
\ \BBA {} Baskaran%
}{%
Stathopoulos%
\ \BBA {} Baskaran%
}{%
{\protect \APACyear {1996}}%
}]{%
stath}
\APACinsertmetastar {%
stath}%
\begin{APACrefauthors}%
Stathopoulos, T.%
\BCBT {}\ \BBA {} Baskaran, B.%
\end{APACrefauthors}%
\unskip\
\newblock
\APACrefYearMonthDay{1996}{}{}.
\newblock
{\BBOQ}\APACrefatitle {Computer simulation of wind environmental conditions
  around buildings} {Computer simulation of wind environmental conditions
  around buildings}.{\BBCQ}
\newblock
\APACjournalVolNumPages{Engineering structures}{18}{11}{876--885}.
\PrintBackRefs{\CurrentBib}

\bibitem [\protect \citeauthoryear {%
Summers%
, Hanson%
\BCBL {}\ \BBA {} Wilson%
}{%
Summers%
\ \protect \BOthers {.}}{%
{\protect \APACyear {1986}}%
}]{%
sum86}
\APACinsertmetastar {%
sum86}%
\begin{APACrefauthors}%
Summers, D.%
, Hanson, T.%
\BCBL {}\ \BBA {} Wilson, C.%
\end{APACrefauthors}%
\unskip\
\newblock
\APACrefYearMonthDay{1986}{}{}.
\newblock
{\BBOQ}\APACrefatitle {Validation of a computer simulation of wind flow over a
  building model} {Validation of a computer simulation of wind flow over a
  building model}.{\BBCQ}
\newblock
\APACjournalVolNumPages{Building and Environment}{21}{2}{97--111}.
\PrintBackRefs{\CurrentBib}

\bibitem [\protect \citeauthoryear {%
Swaddiwudhipong%
\ \BBA {} Khan%
}{%
Swaddiwudhipong%
\ \BBA {} Khan%
}{%
{\protect \APACyear {2002}}%
}]{%
swad}
\APACinsertmetastar {%
swad}%
\begin{APACrefauthors}%
Swaddiwudhipong, S.%
\BCBT {}\ \BBA {} Khan, M.%
\end{APACrefauthors}%
\unskip\
\newblock
\APACrefYearMonthDay{2002}{}{}.
\newblock
{\BBOQ}\APACrefatitle {Dynamic response of wind-excited building using CFD}
  {Dynamic response of wind-excited building using cfd}.{\BBCQ}
\newblock
\APACjournalVolNumPages{Journal of sound and vibration}{253}{4}{735--754}.
\PrintBackRefs{\CurrentBib}

\bibitem [\protect \citeauthoryear {%
Tsuchiya%
, Murakami%
, Mochida%
, Kondo%
\BCBL {}\ \BBA {} Ishida%
}{%
Tsuchiya%
\ \protect \BOthers {.}}{%
{\protect \APACyear {1997}}%
}]{%
tsu97}
\APACinsertmetastar {%
tsu97}%
\begin{APACrefauthors}%
Tsuchiya, M.%
, Murakami, S.%
, Mochida, A.%
, Kondo, K.%
\BCBL {}\ \BBA {} Ishida, Y.%
\end{APACrefauthors}%
\unskip\
\newblock
\APACrefYearMonthDay{1997}{}{}.
\newblock
{\BBOQ}\APACrefatitle {Development of a new k- $\varepsilon$ model for flow and
  pressure fields around bluff body} {Development of a new k- $\varepsilon$
  model for flow and pressure fields around bluff body}.{\BBCQ}
\newblock
\APACjournalVolNumPages{Journal of Wind Engineering and Industrial
  Aerodynamics}{67}{}{169--182}.
\PrintBackRefs{\CurrentBib}

\bibitem [\protect \citeauthoryear {%
Yuan%
, Ng%
\BCBL {}\ \BBA {} Norford%
}{%
Yuan%
\ \protect \BOthers {.}}{%
{\protect \APACyear {2014}}%
}]{%
yuan}
\APACinsertmetastar {%
yuan}%
\begin{APACrefauthors}%
Yuan, C.%
, Ng, E.%
\BCBL {}\ \BBA {} Norford, L\BPBI K.%
\end{APACrefauthors}%
\unskip\
\newblock
\APACrefYearMonthDay{2014}{}{}.
\newblock
{\BBOQ}\APACrefatitle {Improving air quality in high-density cities by
  understanding the relationship between air pollutant dispersion and urban
  morphologies} {Improving air quality in high-density cities by understanding
  the relationship between air pollutant dispersion and urban
  morphologies}.{\BBCQ}
\newblock
\APACjournalVolNumPages{Building and Environment}{71}{}{245--258}.
\PrintBackRefs{\CurrentBib}

\end{thebibliography}
\end{document}